\documentclass[11pt]{article}
\usepackage[round]{natbib}
\usepackage{amsmath}
\usepackage{amsthm}
\usepackage{amssymb}
\usepackage{amsthm}
\usepackage{nccmath}
\usepackage{graphicx}
\usepackage{sidecap}
\usepackage{subcaption}
\usepackage{caption}
\usepackage{comment}
\usepackage{float}
\usepackage{tikz}
\usepackage{pgf, pgfplots}
\usepackage{xcolor}
\usepackage{array}
\usepackage{setspace}
\usepackage{fullpage}
\usepackage{longtable}
\usepackage{pdfpages}
\usepackage{import}
\usepackage{rotating}
\usepackage[hidelinks]{hyperref} 

\usepackage{cleveref} 
\usepackage{minitoc}
\usepackage{xfrac} 

\newcommand{\monthyeardate}{\ifcase \month \or January\or February\or March\or April\or May \or June\or July\or August\or September\or October\or November\or December\fi, \number \year} 
       
\newtheorem*{thm*}{Theorem}
\newtheorem{prop}{Proposition}
\newtheorem*{prop*}{Proposition} 
     
\newtheorem{lem}{Lemma}
\newtheorem*{lem*}{Lemma}
\newtheorem{defn}{Definition}
\newtheorem{rem}{Remark} 
\newtheorem*{rem*}{Remark} 
 
\newtheorem*{ex*}{Example}

\newtheorem{cond}{Condition}
\usepackage{comment}
\definecolor{green}{HTML}{51A351}
\definecolor{blue}{HTML}{2F96b4}
\definecolor{gray}{HTML}{BCBCBC}

\DeclareMathOperator*{\plim}{plim}

\title{More connection, less community: network formation and local public goods provision}
\author{Alastair Langtry\footnote{University of Bristol.  Email: \emph{alastair.langtry@bristol.ac.uk.} I am grateful to Marina Agranov, Toke Aidt, Sebastian Bervoets, Renaud Boules, Yann Bramoulle, Krishna Dasaratha, Christian Ghiglino, Sanjeev Goyal, Gilat Levy, Meg Meyer, Pau Milan, Bryony Reich, Weilong Zhang, Yifan Zhang, and many seminar participants for helpful comments. I am especially grateful to Matt Elliott and DJ Thornton for insightful discussions. Any errors are my own. Declaration of interests: none.}}

\date{\monthyeardate} 
\begin{document}
\maketitle
\begin{abstract}
This paper presents a model of network formation and public goods provision in local communities. Here, networks can sustain public good provision by spreading information about people's behaviour. I find a critical threshold in network connectedness at which public good provision drops sharply, even though agents are highly heterogeneous. Technology change can tear a community's social fabric by pushing high-skilled workers to withdraw from their local community. This can help explain rising resentment toward perceived ``elites'' -- their withdrawal actively harms those left behind. Moreover, well-meaning policies that upskill workers can make them worse off by reducing network connectedness.


\end{abstract}

\maketitle
\newpage

In recent decades, technological progress has made the world a smaller and smaller place. It is easier than ever before to connect with other people -- whether on the other side of the street or the other side of the world. Yet across the Western world, some communities seem to have become more \emph{dis}connected and have seen their social fabric fray. There is a concern that these communities are no longer providing local public goods to their members -- from volunteering and participation in community organisations to keeping streets clean \citep{dietz2019less,forbes2024volunteering, generosity2024}.

Sociological work provides compelling stories of the fraying of some of these communities (see for example, \cite{putnam2016our, vance2022hillbilly}). And these stories line up with more quantitative work showing a decline in community life \citep{putnam2000bowling}. In turn, this loss of community has been linked to a rise in `deaths of despair' \citep{case2020deaths} and higher crime \citep{levellingup2022}.



This paper shows how innovation, including that which makes forming connections easier, can cause some communities to become disconnected and tear their social fabric -- leaving them unable to provide public goods to their members. Within a single community, this fall in public good provision can be sudden and stark. So innovation is not merely increasing inequality, but also leaving some people worse off in absolute terms. It also shows how those who benefit from innovation can be responsible for this tearing of the social fabric, precisely \emph{because} they benefit. Within a community, the winners from the march of technological progress \emph{create} the losers. In a similar vein, providing skills to allow more people to benefit from innovation can actively harm those who are left without those skills. The effects will also be uneven across time: sometimes innovation benefits people, and sometimes it makes them worse off by tearing the social fabric.

The key driving force in the model is that people face incentives to spend time exploiting new technologies -- which crowds out time spent interacting within their local community. This force will be present in technological developments that make connecting with people easier, so long as they make connecting outside of the local community relatively easier. In turn, this `thins out' the social network within a community. But here, the social network plays an important role in incentivising public good provision. It spreads information about people's behaviour, which allows the community to reward its members for their contributions. If the social network becomes too `thinned out', then the information spreading mechanism breaks down. And public good provision drops sharply. We can think of this as the social fabric tearing. 
As new technologies push people to spend time away from the community, this stretches out the social fabric until it tears completely.

In the model, agents are embedded in a local community -- a group of people who know one another's identities and have a shared interest in some local public goods.\footnote{Local neighbourhoods are an enduring source of community for many: at age 30, 80\% of non-graduates in the UK live in the same travel-to-work-area they did at age 16. For university graduates, the figure is still 60\% \citep{DFE_2021}. The US is similar: at age 26, 30\% live in the same census tract at age 16, and 58\% live within 10 miles \citep[S4]{sprung2022radius}. A census tract contains around 4,000 individuals \citep{US_census_tract_2023}. In San Salvador, El Salvador 77\% of people have lived in their neighbourhood for their \emph{entire} life \citep[S4.3, p.20]{melnikov2020gangs}. The form that public goods take will depend on the community in question. It could include keeping the streets clean, campaigning for better children's play areas, offering informal careers advice to young people, or supporting a local school. } 
First, agents choose how much time to spend interacting inside their community, and how much time to spend outside of it. The amount of time someone spends inside their community influences how many friends they have. An overall social network is then formed through a random networks process. Second, agents choose how much to contribute to local public goods. Their action is observed by some others in the community. Information about their action then flows through the social network. Finally, are then some others who provide a social reward -- but only do so if they learn of the contribution. 

It is this social reward that incentivises public good provision. But it relies critically on information transmission through the social network. So here, communities rely on social networks to facilitate public goods. This is in line with the sociological view that networks are or create social capital \citep{coleman1988social, putnam2000bowling}, and that they encourage good behaviour by creating reputations \citep{simpson2015beyond}.

Strikingly, I find that there is a critical threshold in the connectivity of the social network within a community that determines whether people provide public goods. 
This is in spite of heterogeneity in people's cost of providing public goods, systematic differences in people's expected network positions and different people having opportunities to invest in different local public goods. Above the threshold, information about people's behaviour spread widely. This allows the community to reward good behaviour and get high local public good provision. Below the threshold, there is no such accountability, and so little public good provision. The key driver of the result is a large change in how far information spreads when network connectivity crosses the critical threshold.

This means that small changes in technology, creating small changes in the incentives to spend time inside the community, can have large effects on public good provision. In other words, small changes can tear the social fabric. This works by pushing network connectivity across the critical threshold. So an improvement in technology -- one that would make everyone better off were behaviour held constant -- can tear a community's social fabric and make all of its members worse off. 
While a given change in technology pushes network connectivity down in all communities, it may only push it across the critical threshold in some communities. So some communities win, while others lose.

Further, not everyone has the same opportunities to benefit from new technologies. There is substantial evidence that much of the innovation over recent decades has been skill-biased: benefitting skilled workers and managers more than unskilled workers \citep{acemoglu2002technical}. Precisely because they benefit more from new technologies, skilled workers will reduce the amount of time they spend interacting inside their community by more. When this causes the social fabric to tear, unskilled workers will be left worse off by the loss of public goods. So while technological change is ultimately driving the tearing of social fabric, there is a meaningful sense in which skilled workers are responsible for it. This can help ground antipathy felt by parts of some communities towards an `elite' in real economic grievance, not just cultural enmity or misplaced anger.

Finally, attempts to improve the lot of unskilled workers by equipping them with the skills needed to benefit from new technologies can have collateral damage. Those who become skilled workers will spend less time inside their community, as they spend time exploiting new technologies. This reduces network connectivity -- which could tear the community's social fabric. If it does, then those who were left without skills are made worse off. They are not just left behind, but are collateral damage.

\subsection*{Related Literature}
There is a substantial literature in economics and sociology examining role that interactions between people play in community life and public goods provision inside communities (for example, \cite{becker1974theory,bardsley2000interpersonal} and \cite{bowles2002social}). Much of this work either abstracts from an explicit network structure \citep{putnam1995tuning, putnam2000bowling, ahn2008social} or focuses on the importance of common friends \citep{coleman1988social, jackson2012social}. In contrast, my paper highlights the importance of overall network connectedness. Understanding what can drive a decline in public good provision (often phrased as a decline in social capital) has been an important concern in this literature, although the exact extent of the decline remains contested \citep{costa2001understanding, stolle2005inaccurate}. Here, my model contributes by showing how small changes in network structure can have large effects on public good provision, even when agents are highly heterogeneous. And also that any changes can be specific to individual communities. 

By explicitly modelling network formation, my model is able to show precisely how seemingly beneficial technological and social changes can harm welfare by damaging public good provision within communities. This is similar in spirit to \cite{acemoglu2024employment}, who show how technology changes that ought to benefit managers at the expense of workers can make everyone worse off through the way they alter community cooperation.\footnote{A review of the large literature on the harmful effects of innovation is beyond the scope of this paper. Some recent contributions here include \cite{autor2022labor, acemoglu2023distorted, johnson2023power}.} 
My result is starker: innovations that seemingly should benefit everyone in a community can make everyone worse off through the way they damage public good provision within the community.

The network plays a prominent role in my model, and the paper relates to several strands of the networks literature in economics. First, there is work on the private provision of public goods in fixed networks \citep{bramoulle2007public, allouch2015private, elliott2019network}. Second, there is work on cooperation in networks with repeated interactions (usually playing a Prisoner's Dilemma with neighbours). Early contributions here typically consider patient players \citep{kandori1992social,ellison1994cooperation}, with later contributions considering impatient agents \citep{ali2013enforcing,ali2016ostracism, ali2022communication}.\footnote{Other important contributions to this literature include \cite{vega2006building, lippert2011networks, nava2014efficiency} and \cite{jackson2012social}.} 
My setting differs from these strands in two important ways. First, public good provision/cooperation is with the community as a whole, rather than just bilaterally with neighbours. Second, my mechanism is based on information spreading. It does not rely on strategic inter-dependencies in neighbours' actions (as in the public goods strand) or on repeated interactions (as in the cooperation strand) to sustain public good provision. 


Third, there is work using random graphs to study diffusion processes (including, for example, \cite{watts2002simple, campbell2013word, campbell2024network}).\footnote{There is also work on diffusion using fixed networks, but my paper is less closely related to this. Early contributions here include \cite{granovetter1978threshold, blume1993statistical, morris2000contagion}. Recent contributions include \cite{langtry2024network, matouschek2025organizing}.}  
In my model, it is the diffusion of information that is important. Compared to existing work, it shows a new way of obtaining a sharp discontinuity in outcomes at a threshold in network connectivity -- by leveraging the emergence of a `giant component'. It is well-understood that the presence of a `giant component' affects economic behaviour in a range of settings; see for example \cite{akbarpour2018diffusion,sadler2020diffusion} and \cite{dasaratha2023innovation}. But most existing work does not show a sharp discontinuity, and has typically relied on interacting two different networks \citep{buldyrev2010catastrophic}, or on having several different types of links \citep{elliott2022supply, elliott2023corporate}. 

Within this literature, \cite{campbell2024strategic} is perhaps closest to my paper. They study a world where people can only contribute to the public good (or public bad) if one of their friends has done so. Essentially, people only learn that an action is available if they see a friend do it. In this setting, actions and information must `travel' together. This is what creates the incentive to contribute to the public good (or refrain from contributing to the public bad): your action may be how friends learn of its existence, and then take the action themselves. 
They also find a sharp discontinuity in contributions at a critical threshold in network connectivity -- although, interestingly, this is only in the case of public bads. In my model, there is no difference in behaviour for public goods and public bads, and only information -- not actions -- is diffusing through the network.

\paragraph{A Roadmap.} The rest of the paper is organised as follows. \Cref{sec:model} sets out the model. \Cref{sec:analysis} characterises equilibrium behaviour. \Cref{sec:welfare} examines the impact of innovation, and \Cref{sec:upskilling} examines the impact of upskilling parts of the community. \Cref{sec:interpretation} provides interpretation of some of the model's important features. \Cref{sec:generalisation} generalises and it in some key directions. \Cref{sec:conclusion} concludes. All proofs are deferred to \Cref{sec:proofs}.

\section{Model}\label{sec:model}
I consider a sequence of games, $\{ \Gamma^{(n)} \}_{n \in \mathbb{N}}$, and set out the game for a given value of $n$. For clarity, I omit the superscript `$(n)$', which formally should appear on all agent-specific objects.

\paragraph{Agents, Actions, Timing.} There are $n$ agents, $i \in \{1, ... ,n\}$, who make up a \emph{community}. A fraction $f \in (0,1)$ of agents are \emph{high-skilled}, and a fraction $(1-f)$ are \emph{low-skilled}. In an abuse of notation, let $H$ and $L$, respectively, denote typical agents from each group and the set of agents in each group. 
The game has two stages. In stage 1, all agents simultaneously choose an amount of time interacting inside their community, $t_i \in [0,1]$, and outside their community, $t_i^{out} \in [0,1]$. At the end of stage 1, Nature moves, forming a network. I describe this process below. In stage 2, all agents simultaneously choose how much to contribute to a public good, $x_i \in [0,1]$. At the end of stage 2, information about agents' contributions spreads through the network and agents receive rewards for contributions. I describe this formally below.

\paragraph{Network formation.} The number of friends an agents has (her \emph{degree}), $d_i$, depends on how much time she spends interacting inside her community. Formally, $d_i$ is a random draw from a Probability Mass Function $p(d | t_i) = P(d_i = d | t_i)$ that does not depend on $n$. I assume that: (1) the expected degree is increasing in $t_i$, (2) there is some maximum possible degree that is independent of $n$, and (3) there is always some positive probability of having degree one.  Additionally, I impose a technical assumption that: (4) $\mathbb{E}[d^2 | t] - 2 \mathbb{E}[d | t]$ is strictly increasing in $t$, with $\mathbb{E}[d^2 | t] - 2 \mathbb{E}[d | t] = 0$ for some $t \in (0,1)$.\footnote{Formally, (1) $\mathbb{E}[d | t]$ is strictly increasing in $t$, (2) $P(d_i = 1 | t_i) > 0$ for all $t_i \in [0,1]$, and (3) there exists some $D > 3$, such that for all $d > D$; $P(d_i = d | t_i) = 0$ for all $t_i \in [0,1]$. (4) ensures that network connectivity, in the sense that will matter in this setting, is increasing in time inside the community. \Cref{OA:tech_ass} discusses this assumption in detail. Finally, if $\sum_i d_i$ is initially odd, then randomly select an agent with non-zero degree, and reduce her degree by one. This ensures that $\sum_i d_i$ is even.} 
The set of resulting draws, $\mathbf{d}^{(n)} = (d_j)_{j=1}^n$, is the \emph{degree sequence}. 

For each agent $i$, Nature then draws an unweighted and undirected network $G(i)$ uniformly at random from those with the degree sequence $\mathbf{d}_{-i}^{(n)} = (d_j)_{j \neq i}^n$. Each network $G(i)$ is drawn independently, and does not include $i$.


\paragraph{Information spreading and rewards.} At the end of stage 2, a randomly chosen set of agents, denoted $\mathcal{N}_i^o$, `observe' $i$'s action. These \emph{observers} tell their friends about it, who tell their friends, and so on. This communication is non-strategic. So for every $j \in \mathcal{N}_i^o$, all agents in the same connected component as $j$ learn of $i$'s action. A randomly chosen set of agents, denoted $\mathcal{N}_i^r$, can `reward' $i$'s action; but only if they learn of $i$'s action. Among the \emph{rewarders} who learn of $i$'s action (if any), one is picked at random and provides a reward non-strategically.\footnote{This takes the view that information spreading is incidental: agents make friends for other reasons, and just talk about other people's actions. Similarly, rewards are incidental or free to provide. \Cref{sec:interpretation} how to interpret rewards in more detail, and \Cref{OA:gen_info} considers a setting where both communication and reward provision are strategic decisions.}

I assume that 
the number of observers and the number of rewarders are both on the order of $n^{\alpha}$ (denoted $\mathcal{O}(n^{\alpha})$), for some $\alpha \in (0, \frac{1}{6})$.\footnote{Formally, there exist positive constants $k_1, k_2 > 0$ and $n_0$ such that for all $n > n_0$, we have $k_1 n^\alpha \leq |\mathcal{N}_i^o| \leq k_2 n^\alpha$ and $k_1 n^\alpha \leq |\mathcal{N}_i^r| \leq k_2 n^\alpha$ for all $i$.} 
Intuitively, this requires that when the community becomes large, agents have `many' people who observe their action and can reward them, but the fraction of such people is vanishingly small.

\paragraph{Preferences.} Each agent receives some  direct benefits to spending time inside and outside of her community, with diminishing returns to each use. The returns to spending time outside of her community depend on her group: high-skilled agents have a higher return than low-skilled agents (both in absolute terms and at the margin).
She also faces a convex cost of total time (implicitly, an opportunity cost of spending time alone; e.g., sleeping). Further, she has some constant marginal cost of contributing to public goods, and receives some benefit proportional to her contribution \emph{if and only if} a reward is provided to her. Finally, she benefits when others contribute to public goods. So $i$'s preferences are captured by:

\begin{align}\label{eq:prefs}
    u_i = I(t_i, \pi_i t_i^{out}) - c(t_i + t_i^{out}) +  p_i(\text{reward}) R x_i - C_i x_i + B\left( \mathbf{x_{-i}} \right),
\end{align}

where $I(\cdot, \cdot)$ is strictly increasing and concave in both arguments, $c(\cdot)$ is strictly increasing and convex with $\lim_{x \to 1} c'(x) = \infty$, $B(\cdot)$ is strictly increasing. For convenience, assume that all functions are twice continuously differentiable. Also, $R>0$, $\mathbf{x_{-i}} = \frac{1}{n-1} \sum_{j\neq i} x_j$, $\pi_i = \pi_H$ for high-skilled agents, and $\pi_i = \pi_L$ for low-skilled agents, with $\pi_H > \pi_L > 0$, and $C_i \in \mathbb{R}$ is a random draw from some Probability Mass Function that has finite support and does not depend on $n$, with associated Cumulative Distribution Function $F(C)$. 
Finally, $p_i(\text{reward})$ is the probability that $i$ receives a reward for her contribution to public goods.


\paragraph{Information.} Each agent knows their own utility function, how the network forms and how information spreads. Agents' identities (i.e. their index $i$) are also common knowledge. After Nature forms the network, but before the start of stage 2, all agents learn their own degree, $d_i$, and the overall degree sequence of the network.\footnote{Common knowledge is not needed here, because there are no strategic interactions here. Further, agents do not need to learn the realisation of the network (although nothing would change if they did). This is a weaker assumption than the standard in much of the networks literature -- namely that agents know the whole network.} 

\paragraph{Solution Concept.} I will focus on Nash Equilibria in which agents' choices of how much time to spend inside their community converge to some distribution as the number of agents becomes large. Sequences of Nash Equilibria that never converge are less appealing because behaviour remains forever sensitive to the addition of more agents. 

\begin{defn}[Convergent Nash Equilibrium]\label{defn:convergent_NE}
    a sequence of action profiles $\{ (t_i^*, t_i^{out *}, x_i^*)_{i=1}^{n} \}_{n \in \mathbb{N} }$ such that: (1) for each $n$, $(t_i^*, t_i^{out *}, x_i^*)_{i=1}^{n}$ is a Nash equilibrium, and (2) $\Lambda^{(n)}(t^*) \overset{d}{\to} \Lambda(t)$,
\end{defn}

From a technical standpoint, this convergence in first stage play is important for characterising second stage behaviour. And in turn, a characterisation of second stage behaviour is important for characterising first stage behaviour. With this in place, I now turn immediately to characterising behaviour, and defer discussion of 
how to interpret some features of the model to \Cref{sec:interpretation} and of possible generalisations to \Cref{sec:generalisation}. 


\section{Behaviour in the Community}\label{sec:analysis}
\subsection{Contributing to Public Goods}
I begin by characterising second stage behaviour: how people contribute to public goods in their community. Given the setup -- that agents find contributions costly but get a reward if the network spreads information about their contribution to someone who can reward them -- the focus is naturally on how network structure shapes these contributions. It turns out that behaviour is governed by a critical threshold in overall network connectivity. Below this threshold, only people with $C_i \leq 0$ will contribute -- i.e. those who do not find contributing costly. Above the threshold, everyone with a cost $C_i < R$ will contribute. Those with $C_i>R$ would never contribute under any circumstances, so it is not surprising that network structure does not affect their behaviour. 

Strikingly, the behaviour of all agents with $C_i \in (0, R)$ switch at the same threshold, even though they are highly heterogeneous in terms of how costly they find public good provision and their degree in the social network.\footnote{They may also differ in their first stage choices, and in the number of observers and rewarders they have.} 
Despite their heterogeneity, a small change in network connectivity at the critical threshold creates a large discontinuous change in the incentives that agents face.

To state this result formally, we will need to assume that the degree sequence `settles down' to some distribution as the number of agents in the community becomes large. Formally, the sequence of degree distributions, $\{ \mathbf{d}^{(n)} \}_{n \in \mathbb{N}}$, is such that $\lim_{n \to \infty} \frac{1}{n} |\{i : d_i = d\}| = \lambda_d \geq 0$ for all $d$. Having convergent first stage choices -- in the sense of \Cref{defn:convergent_NE} -- is necessary and sufficient for this assumption to hold. 
So assuming it here amounts to a focus on Convergent Nash Equilibrium.

\begin{prop}\label{prop:second_stage}
    Suppose $\{ \mathbf{d}^{(n)} \}_{n \in \mathbb{N}}$, is such that $\lim_{n \to \infty} \frac{1}{n} |\{i : d_i = d\}| = \lambda_d \geq 0$ for all $d$. Then there exists an $n_0$ such that for all $n > n_0$,
    
    (i) \ if $\sum_{d} d(d-2) \lambda_d > 0$, then $x_i^* = 1$ if $C_i \leq 0$ and $x_i^* = 0$ otherwise,
    
    (ii) if $\sum_{d} d(d-2) \lambda_d \leq 0$, then $x_i^* = 1$ if $C_i < R$ and $x_i^* = 0$ otherwise.
\end{prop}

This result shows how the a community's social fabric -- the network structure that facilitates public good contributions -- can not just fray, but tear suddenly.\footnote{While I use the term `social fabric' here, it matches the definition of social capital adopted in influential work by \cite{putnam1995tuning, putnam2000bowling}, \cite{coleman1988social} and \cite{jackson2020typology}. Note however, that social capital has been defined in a number of different ways \citep{dasgupta2000social}. For example, \cite{glaeser2002economic} defines it as akin to a version of physical capital, \cite{guiso2006does,guiso2011civic} define it as values and beliefs, and \cite{bourdieu1986forms} views it as power relations -- although this latter view has gained less traction within economics.} 
If overall network connectivity drops below a threshold, public good contributions drop sharply. 

Exactly what these public goods are will depend on the community, and should be viewed broadly. It may be lending food and fuel \citep{jackson2012social}, support with childcare \citep{desmond2012disposable, desmond2016evicted}, volunteering in community organisations, abiding by (positive) social norms, refraining from socially harmful behaviour, or a more nebulous `pride in place' \citep{tanner2020state}. The exact action that constitutes public good provision can also vary across agents. The tearing of social fabric can therefore be felt differently in different communities.


In my model, the way the network facilitates public good contributions is through the way it spreads information. When it reliably spreads information about people's contributions to those in a position to provide a reward, it will sustain pubic good contributions. The key intuition behind the result is that this information spreading mechanism breaks down when network connectivity falls below the threshold. Above the threshold, the network is sufficiently connected and information spreads `widely'. To be more precise, it spreads to a positive fraction of all agents, even as $n$ gets large. So information about $i$'s reliably reaches people who can provide a reward, and hence hold her accountable for her behaviour. This provides incentives for her public good contributions. 

But below the threshold, the same information spreads to a vanishing fraction of all agents as $n$ gets large. This is because the network lacks any `large' components. So the information will not reach people who can provide $i$ with a reward. So she lacks an incentive to contribute. The sudden change in level of contributions is driven by a sudden change in how widely information spreads.

\Cref{prop:second_stage} also is precise about the measure of `network connectivity' that matters: it is exactly $\sum_d d(d-2) \lambda_d$. When I say `network connectivity' throughout the paper, this is the object I am referring to. The measure is closely related to the average degree of the network, but gives more weight to higher degrees.\footnote{Interestingly, agents with degree $1$ make a negative contribution to this measure. It is exactly this feature that makes assumption (4) in the `Network formation' part of \Cref{sec:model} necessary. The intuition behind this feature is somewhat subtle: \Cref{OA:tech_ass} provides a careful discussion.}

\paragraph{Sketching the proof to Proposition \ref{prop:second_stage}.} It is clear from the set-up that an agent contributes if and only if $p_i(\text{reward}) R \geq C_i$ (i.e. the expected benefit outweighs the cost). The probability she gets a reward is then the probability that information about her action passes through the network from an observer to at least one rewarder. Because information is flows `freely', this is simply the probability that any observer shares a connected component with at least one rewarder. Recall that observers and rewarders are spread randomly through the network, and the number of each is growing in $n$. 

So if there is a `giant component' (a component that is on the order of $n$), the probability that at least some of each are in this giant component approaches 1 as $n$ becomes large. If there is not a giant component, then the largest component at most on the order of $n^{\sfrac{2}{3}}$.\footnote{This is a known result regarding random networks. See for example, \cite{janson2009new}. If a component is `on the order of $n^{\sfrac{2}{3}}$' (often written $\mathcal{O}(n^{\sfrac{2}{3}})$), then there exists $k_2 > 0$ and $n_0 > 0$ such that for all $n > n_0$ we have $|\mathcal{C}| \leq k_2 n^{\sfrac{2}{3}}$, where $\mathcal{C}$ is the set of agents in the component.} 
So the fraction of agents in this largest component vanishes as $n$ becomes large. 
%
As an important technical detail, it goes to zero `faster' than the number of `rewarders' goes to infinity. So even in the best case scenario (in terms of $i$ being able to get a reward), the probability that a `rewarder' learns about $i$'s action goes to zero as $n$ becomes large.
The final step is to apply existing results from the graph theory literature which show that the giant component exists if and only if network connectivity passes a critical threshold, and characterises that threshold. 

\subsection{Spending Time}
Now I characterise first stage behaviour: how people spend their time. The headline result is simple. When there are a large number of agents in the community, the impact that any one agent has on the second stage vanishes. This is because only overall network connectivity that matters for second stage behaviour, and one agent cannot affect it. So they simply follow their `direct' incentives to spend time.


\begin{prop}\label{prop:first_stage}
There exists a convergent Nash equilibrium. In all convergent Nash equilibria, there exists an $n_0$ such that for all $n > n_0$: $t_i^* = \hat{t}_i$ and $t_i^{out *} = \hat{t}^{out}_i$ for all $i$, where $(\hat{t}_i, \hat{t}^{out}_i) = \arg\max I(t_i, \pi_i t_i^{out}) - c(t_i + t_i^{out})$.
\end{prop}

This means that any factors affecting first-stage incentives will feed through very cleanly into overall network connectivity, and then through into public good provision within the community. This will help with the comparative statics exercises that are the focus of the next section. But first, I will briefly comment on the proof. While the intuition behind the result is very simple, proving it in a random networks setting requires a few steps. 

\paragraph{Sketching the proof to Proposition \ref{prop:first_stage}.} First, I show that if first stage behaviour is convergent (in the sense of \Cref{defn:convergent_NE}) then the assumption in \Cref{prop:second_stage} is satisfied. This uses Kolmogorov's Strong Law of Large Numbers to account for the fact that agents can have different choices of $t_i$, and hence that their degrees are drawn from different distributions. It is important because we need a characterisation of second stage behaviour to analyse first stage behaviour. 

Second, it shows that changing one agent's degree has no impact on the probability that a giant component exists, and hence no impact on second stage behaviour for sufficiently large $n$. This is because one agent, with a degree that only vary between $0$ and some fixed $D$, can only break up a component into $D$ pieces or combine together $D$ different components. So then agents will simply choose $t_i, t_i^{out}$ to maximise their first stage payoffs. Finally, this first stage behaviour is convergent -- which is required for the characterisation to be valid.\footnote{The exact equality ($t_i^* = \hat{t}_i$) is due to there being finitely many different values that $C_i$ can take. So there is a `gap' in the distribution of costs between zero and the smallest strictly positive value (and similarly between $R$ and $\max\{C : C<R\}$. If $C_i$ were drawn from a continuous distribution, we would have $t_i^* \to \hat{t}_i$ and $t_i^{out *} \to \hat{t}_i^{out}$ as $n \to \infty$. The online appendix provides technical discussion of this.}

\subsection{Discussion}\label{sec:policy_discussion}
So far, we have seen that public good provision depends critically on whether people collectively spend enough time inside their community, but that their choice of where to spend time is shaped by other factors. This allows us to explain sharp falls in public good provision (such as the sharp fall in social capital documented in Robert Putnam's (\citeyear{putnam2000bowling}) seminal \emph{Bowling Alone}) without needing to appeal to large changes to underlying fundamentals.

It is important to note that existing evidence on social capital and the provision of public goods within communities is typically reported at the aggregate level. My model 
makes a more granular prediction. Within an individual community, public good provision drops sharply when network connectivity falls below a critical threshold. But there is no reason to suppose that these sharp drops should happen at the same time in different communities. So a smooth (if rapid) decline in social capital at the aggregate level is consistent with the sudden collapses in individual communities.

An important exception to this is \cite{chay2012black}, who look at political participation and church membership within Black communities in the US South during the Reconstruction Era (1870-1890).\footnote{Both political participation and church membership are forms of public goods in these communities during the Reconstruction Era. See \cite{chay2012black} for discussion.}
They present empirical evidence of a threshold in network connectedness below which connectedness has no impact on public good provision, and above which it does. Past the threshold, they find a positive relationship between their community connectedness measure and the outcomes of interest.\footnote{\cite{chay2012black} use the planting of labour intensive crops as the county level as a proxy for social connectedness. The key idea is that the density of the black population is higher where there are more labour intensive crops grown, and density is a binding constraint on community connectedness.} 
However, their measure also picks up a community's size, as well as the density of connections. So we should expect this positive relationship past the threshold, even if the rate of participation jumps only at the threshold (as predicted by \Cref{prop:second_stage}).

Another feature of my model is that people can interact with other both inside and outside of their community, but only the former matters for sustaining public good contributions. This means that social, technological, or other changes that make interacting with others easier or more attractive can still tear a community's social fabric. There may be more interactions overall, but fewer of the within-community ones that sustain the social fabric. This is how we can have more connection, but less community.  

\section{The impact of new technology}\label{sec:welfare}
I now examine the impact of innovation, one such change that can make interactions easier or more attractive. Within the model, this is treated as an increase in the parameter $\pi$ for some or all agents. The key idea is that people must spend time using technologies. By improving technologies, innovation increases the incentives to spend time using them. This can cover a wider range of technologies: including both those that people must literally spend time using (for example, new machines or software) and those like institutional innovations that raise the payoffs to using existing technologies. It would also cover an increase in the stock of ideas held outside the community -- which could increase the (relative) benefits of interacting outside of the community by making recombinant innovation more attractive \citep{hargadon2003breakthroughs}.

For technologies that people use for interacting with others, the argument is somewhat different. For these, innovation leads to easier or better interactions, and so raises the payoffs to interacting with others. While this kind of innovation makes interactions of all kinds easier, at least anecdotally, they are skewed towards making it relatively easier to interact with people outside of the local community. For example, new communication, transport, and matching technologies have made it somewhat easier to interact with neighbours across the street. But they have made it vastly easier to interact with those from other communities. So modelling innovations in these technologies as an increase in $\pi$ focuses on changes in \emph{relative} ease of interactions, and abstracts away from an absolute component that applies both inside and outside the community.\footnote{It would be straightforward to capture this with an additional parameter that acts multiplicatively on the function $I(\cdot , \cdot)$. But I abstract away from this for convenience.}

Further, there is substantial evidence that much innovation over recent decades has been skill-biased: benefitting high-skilled more than low-skilled agents \citep{acemoglu2002technical, aghion2002schumpeterian, violante2008skill, buera2022skill}. It has given high-skilled agents greater opportunities to exploit improved technologies. Within my model, this shows up as $\pi_H$ increasing by more than $\pi_L$. For cleanliness, I focus on a benchmark case where innovation increases \emph{only} $\pi_H$ (the qualitative insights are of course not sensitive to this).

In light of \Cref{prop:first_stage}, the comparative static on $\pi_H$ is straightforward. When $\pi_H$ rises, high-skilled agents spend more time outside of their community, and less time inside it. This pushes network connectivity down. In turn, this can tear the community's social fabric -- i.e., cause a sharp drop in public good contributions.

\begin{rem}\label{rem:comp_stat_pi}
    In all convergent Nash equilibria, for sufficiently large $n$: network connectivity, $\sum_d d(d-2) \lambda_d$, is strictly decreasing in $\pi_H$. Public good contributions, $x_i^*$, are weakly decreasing in $\pi_H$.
\end{rem}

An immediate implication is that innovation that benefits only the high-skilled can destroy the social fabric for a whole community. This is because high-skilled agents gradually withdraw from spending time inside their community. That low-skilled agents do not, and continue spending just as much time in their community as before, does not protect them. It is overall network connectivity that matters, not the time spent by any one group. 

\subsection{Welfare} 
When it does not tear a community's social fabric, innovation (increasing $\pi_H$) directly increases payoffs to high-skilled agents. An adjustment in how high-skilled agents spend their time -- less inside their community and more outside of it -- further increases their payoffs. But this adjustment reduces network connectivity. This can tear the social fabric, bringing a sharp drop in public good contributions. This harms everyone in the community -- not just high-skilled agents.

\begin{prop}\label{prop:welfare}
    In all convergent Nash equilibria, for sufficiently large $n$: 
    
    (i) \ for high-skilled agents, utility is strictly increasing in $\pi_H$, except when the increase in $\pi_H$ reduces public good contributions.

    (ii) for low-skilled agents, utility is weakly decreasing in $\pi_H$, and strictly so if the increase in $\pi_H$ reduces public good contributions.
\end{prop}

\begin{figure}[h]
\centering
\begin{tikzpicture}[thick, scale=1]
\draw[->] (0,0) -- (10,0) node[right] {$\pi_H$};
\draw[->] (0,0) -- (0,6) node[left] {$u_i$};
\node at (0.1,-0.4) {$\pi_L$};
\draw[very thick,green,domain=0:5,smooth] plot (\x, {2.5 + 0.2*\x + 0.05*\x*\x}); 
\draw[very thick,dashed,green] (5,4.75) -- (5,3.25);       
\draw[very thick,green,domain=5:10,smooth] plot (\x, {2.2 + 0.2*\x + 0.05*(\x-4)*(\x-4)}); 
\draw[very thick,red,domain=0:5,smooth] plot (\x,2.5); 
\draw[very thick,dashed,red] (5,2.5) -- (5,1);       
\draw[very thick,red,domain=5:10,smooth] plot (\x, 1); 
\end{tikzpicture}
\caption{Green line shows equilibrium payoffs for high-skilled agent. Red line shows payoffs for low-skilled agent.}
\label{fig:pi_H}
\end{figure}
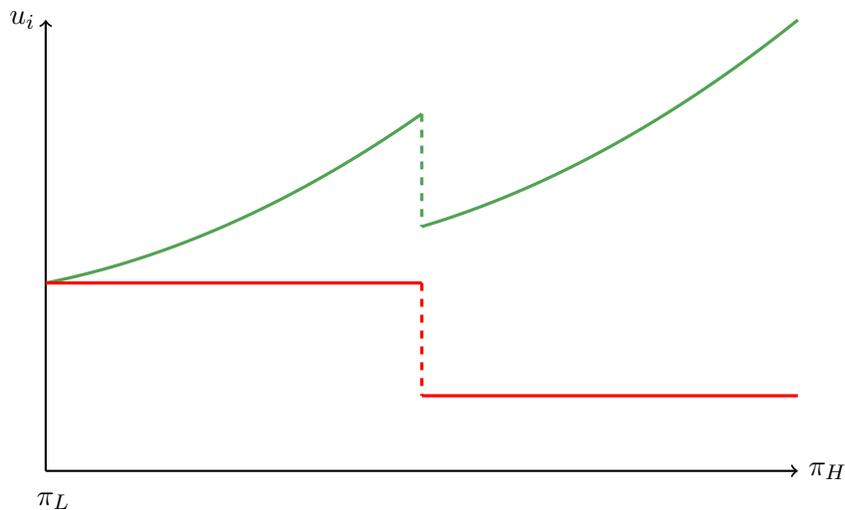

\Cref{prop:welfare} cannot guarantee that innovation definitely tears the social fabric. This is because the model has not ruled out the cases where: (a) low-skilled agents are numerous enough and spend enough time inside the community to sustain public good contributions by themselves, or (b) public goods contributions were low to below with (i.e. there was no functioning social fabric in the first place). But aside from these cases, innovation will eventually tear the social fabric, by pushing high-skilled agents to withdraw from interacting inside their community. \Cref{fig:pi_H} illustrates this interesting case.

In reality, innovation is a continuing process: we expect that $\pi_H$ rises over time. An implication of \Cref{prop:welfare}, even though it is a static result, is that the welfare effects of innovation are uneven over time.\footnote{Formally, one would extend the model to include chronological time, and make $\pi_H$ an increasing function of chronological time. This would be straightforward, but would not provide any additional insight.}
Initially, rising $\pi_H$ benefits high-skilled workers and harms nobody. Then is causes the community's social fabric to tear -- making everyone worse off. Finally, it reverts to benefitting high-skilled agents.
In short, innovation sometimes improves outcomes and sometimes worsens them.\footnote{A limitation of this welfare analysis is that the model has pushed \emph{global} public goods into the background. These are public goods that affect many communities, and may be influenced by the amount of time agents spend interacting between different communities.}

Given this, it is possible that a social planner -- even one who can make lump sum transfers between agents, and so who does not care about inequality -- prefers not to have innovation. This is because the tearing social fabric creates medium-term losses that could outweigh long-run gains. Note that this is only true if the social planner is impatient (i.e. has a discount factor less than 1) and cannot make unconstrained transfers across time.

Further, we could easily have innovations that low-skilled workers can exploit to some extent (i.e. $\pi_L$ increases as well as $\pi_H$, although by less). It is clear that in this case, payoffs are strictly increasing for both groups except when the innovation tears the social fabric -- when it harms both groups.


\subsection{Inequality}
It is clear that innovation increases between-group inequality when public goods contributions are unaffected. It follows from the assumption that only high-skilled agents benefit from innovation. But more interestingly, between-group inequality also rises when innovation causes the social fabric to tear, and public goods contributions to fall. This is because both high-skilled and low-skilled agents lose the same amount when the social fabric tears. But low-skilled agents lose a larger \emph{proportion} of their overall utility. 

\begin{rem}
In all convergent Nash equilibria, for sufficiently large $n$:
    between-group inequality, measured by a Gini coefficient, is strictly increasing in $\pi_H$.
\end{rem}

An important caveat here is that what is lost when the social fabric tears is public good contributions in a community. These are typically difficult to measure, and are usually omitted from the official statistics that go into most measures of inequality. Additionally, this increasing inequality need not always be the case for an individual level measure. This is because the loss of rewards (which in turn leads to the loss of public good contributions) reduces payoffs for those with smaller $C_i$ by more. That is, within a group, those who were better off initially lose out more from the tearing social fabric. So inequality within each group can fall. And this could, in principle, outweigh the increase in between-group inequality.

Here, the equal loss for both groups is an artefact of the additively separable preferences I use (\Cref{eq:prefs}). The result could be overturned if the `richer', high-skilled, agents lose more from the tearing of the social fabric. But perhaps a more plausible alternative is that the `poorer', low-skilled, agents lose more. For example, if volunteer-led after school clubs close because of a drop in the number of volunteers, poorer families may be harmed more as they are less able to substitute for privately provided childcare.\footnote{This follows along similar lines to an example in \citet[Ch.17, p.289-290]{putnam2000bowling}.}



\subsection{Antipathy}
Beyond increasing inequality, innovation -- by causing the social fabric to tear -- makes low-skilled agents worse off in \emph{absolute} terms.\footnote{Or really any changes that swings incentives away from spending time interacting inside the community. This can include shocks with no apparent connection to life in local communities. For example, a global trade shock that changes a country's comparative advantage, in turn leading to changes in the pattern of labour demand that pushes people to take jobs far from their local community.} 
But while innovation is the ultimate cause in my models, there is nevertheless a meaningful sense in which high-skilled agents are \emph{responsible} for tearing the social fabric. It is the behaviour of high-skilled agents -- their gradual withdrawal from their community in response to innovation -- that is the direct cause of tearing social fabric.

Viewed through the lens of my model, antipathy felt towards `elites' in some parts of Western societies is well-founded. There is substantial anecdotal discussion that antipathy towards an `elite' has contributed to major political events of the past decade \citep{bickerton2016elite, galston2018populist}. An important part of this discussion is a supposed `disconnect' between the `elite' and local communities \citep{may2017}. 

My model can explain how `elites' may have become more disconnected from local communities, and why those who have been `left behind' have a real economic grievance against them. It does not rely on the disconnect inducing policy decisions or other actions that then harm low-skilled (left behind) agents. Rather, the disconnect itself can be harmful.
%
%
In light of this, one natural policy response is upskilling -- converting some low-skilled agents into high-skilled agents. I turn to this in the next section.

\section{Upskilling and the left behind.} \label{sec:upskilling}
Providing more people with the skills to benefit from innovation is widely suggested as a desirable policy response technological change \citep{PWC2019skills, OECD2019future, EC2020skills}. Within my model, this kind of upskilling shows up as an increase in the fraction of agents who are high-skilled. Holding the social fabric fixed, this is clearly beneficial.

But changing the composition of a community -- the fraction of high-skilled vs. low-skilled agents -- \emph{can} affect the social fabric. Increasing the fraction of high-skilled agents, for example through education policies or regional development policies, will reduce the amount of time that agents overall spend inside the community. This is because high-skilled agents spend less time inside the community than low-skilled agents. In turn, this can tear the social fabric. 

\begin{prop}\label{prop:comp_stat_f}
    In all convergent Nash equilibria, for sufficiently large $n$: 
    for remaining low-skilled agents, utility is weakly decreasing in $f$, and strictly so if the increase in $f$ reduces public good contributions.
\end{prop}

The logic here is much the same as for \Cref{prop:welfare}. Those who gain skills partially withdraw from their community in order to spend time making use of their newfound skills. So \Cref{prop:comp_stat_f} highlights an unintended consequence of policies that provide some people with the skills needed to benefit from new technologies. These policies can have collateral damage: through the way they indirectly affect connectedness in the community, these policies can actively harm those who they do not reach. 

\begin{figure}[h]
\centering
\begin{tikzpicture}[thick, scale=1]
\draw[->] (0,0) -- (10,0) node[right] {$f$};
\draw[->] (0,0) -- (0,6) node[left] {$u_i$};
\draw[-] (4,-0.15) -- (4,0.15);
\node at (4,-0.4) {$f_0$};
\draw[-] (6,-0.15) -- (6,0.15);
\node at (6,-0.4) {$f_1$};
\draw[very thick,green,domain=0:5,smooth] plot (\x, {4.5}); 
\draw[very thick,dashed,green] (5,4.5) -- (5,3);       
\draw[very thick,green,domain=5:10,smooth] plot (\x, {3}); 
\draw[very thick,red,domain=0:5,smooth] plot (\x,2.5); 
\draw[very thick,dashed,red] (5,2.5) -- (5,1);       
\draw[very thick,red,domain=5:10,smooth] plot (\x, 1); 

\draw[blue, fill=blue] (6,2) circle[radius=0.1];
\draw[red, fill=red] (4,2.5) circle[radius=0.1];
\end{tikzpicture}
\caption{Green line shows equilibrium payoffs for high-skilled agent. Red line shows payoffs for low-skilled agent. Intervention moves $f$ from $f_0$ to $f_1$. Blue dot shows ex-ante expected payoff for (initially) low-skilled agents with intervention. Red dot show ex-ante expected payoff without intervention.}
\label{fig:f}
\end{figure}
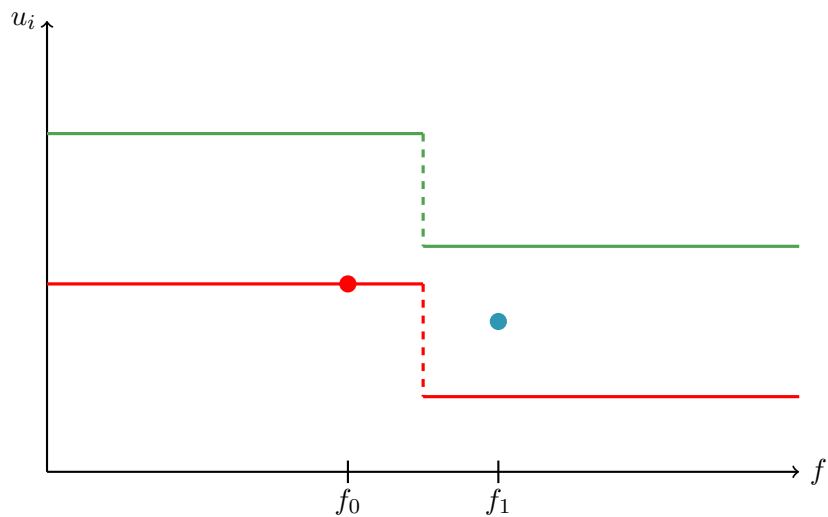

Even more troubling, these interventions can harm all low-skilled agents from an ex-ante perspective. If low-skilled agents are randomly `treated' (i.e. provided with skills), then the expected benefit of the intervention can be negative. This happens when the losses to remaining low-skilled agents from the tearing of the social fabric outweigh the gains to those who are given skills. \Cref{fig:f} illustrates this case: the intervention moves the end fraction of high-skilled agents from $f_0$ to $f_1$. In this example, this causes the social fabric to tear. The red dot in \Cref{fig:f} shows the payoffs of low-skilled agents without the intervention. Then the blue dot shows their ex-ante expected payoff from the intervention.

This suggests that governments should pay attention to the state of a community's social fabric when designing policies. This is true even if the government pays no attention to inequality and only cares about average payoffs. It can also explain why some communities may be resistant to policies designed to give them more opportunities further afield.
To avoid collateral damage, skills policies may benefit from being paired with other interventions designed to facilitate connectedness within a community. 


The remainder of this paper discusses how to interpret some features of the model (\Cref{sec:interpretation}), shows how it can be generalised and extended in a number of directions (\Cref{sec:generalisation}), and finally concludes (\Cref{sec:conclusion}).

\section{Interpreting features of the model}\label{sec:interpretation}
This section provides interpretation of some features of the model: the community, rewards, and friendship formation. This is discussion I deferred from \Cref{sec:model}.

\paragraph{Community.} Two features define a community in my model: (1) agents know one another's identities, and (2) share a common interest in provision of some local public good. The first feature allows the information flow about others' behaviour to be meaningful. Without it, agents can only tell others that someone (unidentified) did or didn't contribute to public goods -- which is not sufficient for people to learn about $i$'s action. The second feature means that agents benefit from public goods provided by others in the community, and gives meaning to the notion of public good provision at the community level. 

This setup provides an exogenous definition of community, distinct from network structure (which is an endogenous object in my model). It also shows how my model could be easily extended to a setting with many communities. Because agents only know the identities of others in their own community, having many communities distinct would have no impact on the outcomes. Each community plays the game independently of all other communities. \Cref{OA:multiple_comms} discusses this in more detail, and shows some of the complications that can arise when people belong to multiple, overlapping, communities at the same time. 

The community can also be viewed as the `social context' in which both the social network and economic behaviour are embedded. The idea that behaviour is embedded in a social context is the default view in anthropology, sociology, and much of the social sciences. When it appears in economics, it often does so in the form of social networks. In contrast, this paper views social networks as themselves being embedded in a social context. This recognises that networks are not the only social context that matter, and also that networks are the result of economic decision-making, 
so themselves might have some social context.

\paragraph{Rewards.} In my model, the reward (1) cannot be provided unless $i$ contributed to the public good \emph{and} some rewarder learned of it, (2) does not grow as more rewarders learn of $i$'s public goods provision, and (3) is costless to provide. One interpretation of rewards is that $i$ can benefit from a favour, and the favour is not costly to provide. Nevertheless, it can only be obtained from someone who knows she is a `good citizen' (and is able to do the favour for $i$); perhaps because people are more inclined to provide favours to `good citizens'.

An alternative view of the reward is that rewarders are people willing to vouch for you -- and that this unlocks some benefits. Agent $j$ can vouch for $i$ only if $j$ knows of $i$'s action. And this only unlocks benefits for $i$ if it reports positive information. Further, $i$ only needs one person to vouch for here -- extra people are of no help. This gives a particular interpretation to the rewarders, $j \in \mathcal{N}_i^r$ -- as the members of the community who know $i$ well enough that they could (in principle) vouch for her, and who have the appropriate social standing/seniority/etc. to allow $i$ to access the associated benefit.

The assumption here is that people provide rewards non-strategically. While I can relax this assumption (see \Cref{OA:gen_info}), there is significant evidence from the sociology literature that people do treat those with a good reputation better -- in terms of respect and cooperation \citep{barclay2004trustworthiness, willer2009groups}, and in market transactions \citep{list2006behavioralist, diekmann2014reputation}, among others -- even absent direct incentives to do so.\footnote{One step further up the chain of reasoning, there is also a body of work in economics that shows people contribute more to public goofs when observed by others \citep{mas2009peers, funk2010social, zhang2011group}. This work focuses on effects coming \emph{directly} from the person observing the action, and does not have a role for the information transmission present in my model and highlighted in the sociology literature.} 

\paragraph{Random networks.} The amount of time people spend inside their community affects the number of friends they have, but not exactly who those friends are: they make friends at random. In this view, meeting others and making friends takes time, but exactly who an agent meets and become friends with is outside of their control and depends on factors that are best modelled as random. See, for example, \citet*{currarini2009economic} and \citet*{cabrales2011social}.\footnote{In both of these frameworks, the number of links formed is a deterministic function of time (or effort) spent building links. This is exactly as in my model. More recent work adopting this convention includes \cite{canen2021endogenous, banerjee2021changes} and \cite{farboodi2023emergence}.} 

This abstracts away from two important aspects of how people form links in practice: (1) that they may intentionally seek out specific people to become friends with, and (2) that common characteristics can create correlations in who is friends with whom. That is, some underlying factor that makes $i$ and $j$ become friends, and $j$ and $\ell$ become friends, may also make it more likely that $i$ and $\ell$ become friends too. 


\paragraph{Public goods and social capital.}
Public good provision is driven by network structure in my model. So the network can be thought of as \emph{social capital}. While the term `social capital' has been defined in a number of different ways \citep{dasgupta2000social}, this view -- that it is the network structures that facilitate and create local public good provision and cooperative behaviour -- is a common one.\footnote{While the `networks' view is perhaps dominant, other conceptions of social capital exist. For example, \cite{glaeser2002economic} define social capital as akin to a version of physical capital, while \cite{guiso2006does,guiso2011civic} define it as values and beliefs and \cite{bourdieu1986forms} views it as power relations -- although this latter view has gained less traction within economics.}  
Notably, it is the view adopted in highly influential work by \cite{putnam1995tuning, putnam2000bowling} and \cite{coleman1988social}. 

\section{Generalising the model}\label{sec:generalisation}
The model set out in \Cref{sec:model} was deliberately kept simple in order to highlight its key features: the tearing of the social fabric when network connectivity falls below a threshold, and the fact that one agent's first stage choice has no impact on the second stage. It generalises and extends in a number of directions. This section discusses four. First, many of the functional form assumptions imposed on the utility function can be relaxed. Second, the information transmission process can be enriched. Third, clustering and homophily -- two features of networks that are ubiquitous in the real-world but which are missing from the random networks model -- can be incorporated straightforwardly. Fourth, it is possible to turn off the second of the key features and give agents some material impact on overall network connectivity. 

\subsection{Utility function}\label{subsec:gen_utility}
The core features that drive behaviour in my model are that: (1) agents get a reward for their public good provision if and only if the connectivity of the overall network is above some threshold, and (2) the choices one agent makes in the first stage do not impact the probability that other agents receive these rewards. This is because any one agent has a vanishing impact on overall network connectivity.
So it is straightforward to generalise the second stage payoffs. Suppose that we have,

\begin{align}
    u_i = I(t_i, \pi_i t_i^{out}) - c(t_i + t_i^{out}) + \Pi_i(x_i , x_{-i} , p_i),
\end{align}
where $\Pi$ is continuously differentiable in all arguments. Suppose the function $BR = [BR_1,...,BR_n]$, where $BR_i = \arg\max_i \Pi_i(x_i , x_{-i} , p_i)$, has a unique fixed point for any given profile $(p_i)_{i=1}^n$.\footnote{Essentially, this assumes that if the second stage were played as a stand-alone game, then it would have a unique Nash equilibrium. See \cite{rosen1965existence, sadler2024simple} for sufficient conditions for a unique equilibrium.} 
And also assume that that the fixed point does not vary with  $(p_i)_{i=1}^n$ when reward probabilities are all sufficiently close to 0 are the same, and similarly when they are sufficiently close to 1.\footnote{Formally, let $x_i^{FX}(\mathbf{p})$ be the unique fixed point of the function $BR: \mathbb{R}^n \to \mathbb{R}^n$. And assume there exists some $k \in (0,1)$ such that $(x_i^{FX}(\mathbf{p}))_{i=1}^n = (x_i^{FX}(\mathbf{p'}))_{i=1}^n$ whenever $p_i, p'_i \leq 1-k$ for all $i$, and that $(x_i^{FX}(\mathbf{p''}))_{i=1}^n = (x_i^{FX}(\mathbf{\hat{p}}))_{i=1}^n$ whenever $p_i, p'_i, p''_i, \hat{p}_i \geq k$ for all $i$.} 
Then the core structure of the equilibrium characterisation will still hold. That is, in all Convergence Nash Equilibria, for sufficiently larger $n$, we will have $t_i^* = \hat{t}_i$ and $t_i^{out *} = \hat{t}^{out}_i$ for all $i$, where $(\hat{t}_i, \hat{t}^{out}_i) = \arg\max I(t_i, \pi_i t_i^{out}) - c(t_i + t_i^{out})$. This is exactly the same as in \Cref{prop:second_stage}. As before, this will induce some level of network connectedness, $\sum_d d(d-2) \lambda_d$. And we will have equilibrium public goods contributions as follows: $x_i^* = x_i^{FX}(\mathbf{p = 1})$ if $\sum_d d(d-2) \lambda_d > 0$ and $x_i^* = x_i^{FX}(\mathbf{p = 0})$ otherwise.

The key point is that the generalised functional form for the second stage payoffs does not change the fact if network connectivity is above the threshold, agents receive a reward for public good contributions with probability tending to 1 (as $n$ becomes large). And conversely, when connectivity is below the threshold, they receive a reward with probability tending to 0 (as $n$ becomes large). For the welfare results to carry through we would additionally need to assume that public goods provision is socially efficient, and that equilibrium contributions are increasing in the probability agents receive a reward for contributing. That is, $\frac{d U^*}{d x_i} > 0$ for all $i$, where $U^* = \frac{1}{n} \sum_{i=1}^n u_i^*$, and $\frac{d x_i^*}{d p_i} \geq 0$, with a strict inequality for some non-empty interval $[p,p'] \subseteq [0,1]$.

\subsection{Information Spreading}\label{subsec:gen_info}
Here, I enrich the information spreading process in two ways. First, by adding frictions but keeping it non-strategic. Second, by making it a strategic decision. I consider each in turn.

\paragraph{Non-strategic frictions.} I generalise the way information spreads in two ways. First, information only flows across a link with some exogenous probability. And second, some fraction of agents never pass on any information. This is a reduced-form way of capturing frictions in information spreading. It reflects the fact that this information is incidental gossip, not the motivating reason for establishing links. So people may not always gossip when they meet. And some people will simply never gossip. From a technical standpoint, adding these frictions amounts to randomly deleting links and nodes respectively from the social network.

This has no impact on the key intuitions, nor on the result that whether or not agents contribute to public goods depends critically on whether network connectivity is above or below some threshold. It only changes the threshold itself. Because links are created at random, removing links at random is essentially the same as not creating them in the first place. So removing links at random is equivalent to adjusting the degree sequence. Similarly, having some agents never pass on information amounts to giving those agents degree zero. What then matters is the network with the \emph{adjusted} degree sequence. \Cref{prop:second_stage} is still driven by the emergence of a giant component -- but now in the adjusted network (with the adjusted degree sequence). \Cref{OA:gen_info} shows this formally.

\paragraph{Strategic information spreading.} The assumption that information flows non-strategically is motivated by evidence that people do spread reputational information about others \citep{feinberg2012virtues, diekmann2014reputation} and provide punishments, even if doing so is costly and there is little or no prospect of repeated interactions in the future \citep{fehr2002altruistic, fehr2007human}. Nevertheless, my model can be expanded to make passing information and providing rewards strategic decisions. 

In \Cref{OA:gen_info}, I make providing a reward and passing information costly actions that agents choose strategically. Crucially, this extension views the provision of rewards and passing of information as themselves a public good that could attract a reward. As with public good provision in the main model, these costs are heterogeneous. Some agents can have a cost that is greater than the potential reward, and others have zero cost. This idea of rewards being provided in order to access future rewards is similar in flavour to work on cooperation in repeated games on networks, such as \cite{jackson2012social}, \cite{nava2014efficiency} and \cite{ali2013enforcing,ali2016ostracism}. 

Unsurprisingly, there can be multiple equilibria in this setting: one where agents both provide rewards and pass information (at least those for whom the reward is less than the cost), and one where they do neither (except for those with zero costs). It is the first such equilibrium that corresponds to the assumption of non-strategic provision. 

\subsection{Real-world network features}\label{subsec:gen_network}
In real social networks, two friends often have (other) friends in common, and those friends are typically similar to one another. This means that real networks typically exhibit significant clustering and homophily. 
However, the configuration models I use produce networks that exhibit no clustering and no homophily (more precisely, as the number of agents becomes large, clustering and homophily shrink to zero). Common friends are vanishingly rare and are no more likely than random to be be of the same type. 

However, it turns out that this does not matter. We could add as much clustering as we want to the model in a very simple way: after Nature forms the network $G(i)$, agents form some additional links with friends of friends. This creates clustering. And if an agent is more likely to form a link with a friend-of-a-friend if they share a type (e.g. their status as a manager or a worker), then this also creates homophily. These links can be added in any way, and we can add as many such links as we like.\footnote{See \cite{jackson2008} and \cite{currarini2009economic}, respectively, for standard definitions of clustering and homophily. It would equally be possible to form links with friends-of-friends-of-friends, or to do multiple rounds of this kind of link formation.} 

This cannot affect \Cref{prop:second_stage}. The reason is simple: any information within a component spreads to all agents in that component. 
Adding links between two agents already in the same component component not change that. And by construction, adding links between friends-of-friends only adds links between two agents who are already in the same component. So it has no effect on who information spreads to -- the extra links are redundant.


An alternative approach would be to use a version of the configuration model that creates random single links \emph{and} random triangles. A version of \Cref{prop:second_stage} would continue to hold because the sudden emergence of the giant component -- which drives my result -- is also present in this alternative version of the configuration model. But the threshold becomes more complicated, and depends on the joint distribution of single links and triangles. See \cite{newman2009random} for details.

\subsection{Impact on second stage.}
The characterisation of the first stage (\Cref{prop:first_stage}) is driven in large part by the fact that how an individual agent spends her time has has no impact on the second stage. This is because there are many agents in the community (i.e. $n$ is large). This feature is needed for traction when working with random graphs. It also provides a useful insight into the case where agents do not consider the impact their first stage decisions have on the overall network connectivity (either because agents really are `small' or because of constraints on cognition). 

A natural alternative is to consider a setting where agents \emph{do} have some non-vanishing impact on overall network connectivity, and hence on the second stage, even if that impact is modest. In terms of the model, this can be done by allowing equal-sized groups of agents to coordinate their choices of $t_i$. This gives agents a non-vanishing impact on overall network connectivity, while maintaining the `large $n$' focus.

The headline outcome is that, whenever public goods would not be provided in the main model \emph{but} provision would a Pareto improvement, this coordination allows the community to sustain public good provision. This is because when connectivity is exactly at the critical threshold, every group's contribution to network connectivity is pivotal. So no group wants to deviate downward and let public good provision collapse. But this new equilibrium is `fragile'. A group's contribution is only pivotal \emph{exactly} at the critical threshold. So public good provision can be destroyed by small unanticipated shocks. \Cref{OA:coord_extension} discusses this in detail.

\section{Conclusion}\label{sec:conclusion}
This paper has examined how communities provide public goods to their own members in a world where having a `good name' matters: that is, where people finding out about how you behaved incentivises good behaviour. I showed that the overall connectedness of the community matters, and that even in highly heterogeneous communities, small changes in connectedness can tear the social fabric -- leading to a large drop in public good contributions. 

Building endogenous network formation on top of this allowed me to examine the impacts of innovation and upskilling. Both seemed beneficial: all else equal, both increased payoffs. But behavioural responses to each of them -- people spending less time inside their community in order to spend more exploiting technological opportunities -- can tear a community's social fabric and harm welfare.

One important simplifying assumption I made in this paper is that people live in one community, and do not move. An interesting avenue for future work could be to study people's choice of which community to belong to, and how belonging to multiple communities at the same time affects outcomes. 

Finally, while this paper is very much focused on life in local communities, the model is more general. At a high level, it examines how network formation and structure affects behaviour in a setting where people can benefit from a reputation for good behaviour. It could be used to study other settings where reputation might matter. One natural application is to behaviour within firms. There are often a range of activities inside firms that cannot be contracted over, and which create significant spillover benefits for the firm as a whole: informal training and mentoring (especially of junior staff) and informal knowledge sharing are two obvious examples. A careful application of my model could shed light on firms' efforts to promote camaraderie and in-person attendance in offices, and help show benefits of the proverbial water cooler.



\newpage
\singlespacing
\bibliographystyle{abbrvnat}
\addcontentsline{toc}{section}{References}
\bibliography{bib}

\begin{thebibliography}{102}
\providecommand{\natexlab}[1]{#1}
\providecommand{\url}[1]{\texttt{#1}}
\expandafter\ifx\csname urlstyle\endcsname\relax
  \providecommand{\doi}[1]{doi: #1}\else
  \providecommand{\doi}{doi: \begingroup \urlstyle{rm}\Url}\fi

\bibitem[Acemoglu(2002)]{acemoglu2002technical}
D.~Acemoglu.
\newblock Technical change, inequality, and the labor market.
\newblock \emph{Journal of economic literature}, 40\penalty0 (1):\penalty0 7--72, 2002.

\bibitem[Acemoglu(2023)]{acemoglu2023distorted}
D.~Acemoglu.
\newblock Distorted innovation: does the market get the direction of technology right?
\newblock In \emph{AEA Papers and Proceedings}, volume 113, pages 1--28. American Economic Association 2014 Broadway, Suite 305, Nashville, TN 37203, 2023.

\bibitem[Acemoglu and Wolitzky(2024)]{acemoglu2024employment}
D.~Acemoglu and A.~Wolitzky.
\newblock Employment and community: Socioeconomic cooperation and its breakdown.
\newblock Technical report, National Bureau of Economic Research, 2024.

\bibitem[Aghion(2002)]{aghion2002schumpeterian}
P.~Aghion.
\newblock Schumpeterian growth theory and the dynamics of income inequality.
\newblock \emph{Econometrica}, 70\penalty0 (3):\penalty0 855--882, 2002.

\bibitem[Ahn and Ostrom(2008)]{ahn2008social}
T.-K. Ahn and E.~Ostrom.
\newblock Social capital and collective action in the handbook of social capital, edited by dario castiglione, jan van deth, and guglielmo wolleb, 2008.

\bibitem[Akbarpour et~al.(2018)Akbarpour, Malladi, and Saberi]{akbarpour2018diffusion}
M.~Akbarpour, S.~Malladi, and A.~Saberi.
\newblock Diffusion, seeding, and the value of network information.
\newblock In \emph{Proceedings of the 2018 ACM Conference on Economics and Computation}, pages 641--641, 2018.

\bibitem[Ali and Miller(2013)]{ali2013enforcing}
S.~N. Ali and D.~A. Miller.
\newblock Enforcing cooperation in networked societies.
\newblock \emph{Unpublished paper.[282]}, 2013.

\bibitem[Ali and Miller(2016)]{ali2016ostracism}
S.~N. Ali and D.~A. Miller.
\newblock Ostracism and forgiveness.
\newblock \emph{American Economic Review}, 106\penalty0 (8):\penalty0 2329--2348, 2016.

\bibitem[Ali and Miller(2022)]{ali2022communication}
S.~N. Ali and D.~A. Miller.
\newblock Communication and cooperation in markets.
\newblock \emph{American Economic Journal: Microeconomics}, 14\penalty0 (4):\penalty0 200--217, 2022.

\bibitem[Allouch(2015)]{allouch2015private}
N.~Allouch.
\newblock On the private provision of public goods on networks.
\newblock \emph{Journal of Economic Theory}, 157:\penalty0 527--552, 2015.

\bibitem[Autor(2022)]{autor2022labor}
D.~Autor.
\newblock The labor market impacts of technological change: From unbridled enthusiasm to qualified optimism to vast uncertainty.
\newblock Technical report, National Bureau of Economic Research, 2022.

\bibitem[Banerjee et~al.(2021)Banerjee, Breza, Chandrasekhar, Duflo, Jackson, and Kinnan]{banerjee2021changes}
A.~Banerjee, E.~Breza, A.~G. Chandrasekhar, E.~Duflo, M.~O. Jackson, and C.~Kinnan.
\newblock Changes in social network structure in response to exposure to formal credit markets.
\newblock Technical report, National Bureau of Economic Research, 2021.

\bibitem[Barclay(2004)]{barclay2004trustworthiness}
P.~Barclay.
\newblock Trustworthiness and competitive altruism can also solve the “tragedy of the commons”.
\newblock \emph{Evolution and Human Behavior}, 25\penalty0 (4):\penalty0 209--220, 2004.

\bibitem[Bardsley(2000)]{bardsley2000interpersonal}
N.~Bardsley.
\newblock Interpersonal interaction and economic theory: the case of public goods.
\newblock \emph{Annals of Public and Cooperative Economics}, 71\penalty0 (2):\penalty0 191--228, 2000.

\bibitem[Becker(1974)]{becker1974theory}
G.~S. Becker.
\newblock A theory of social interactions.
\newblock \emph{Journal of political economy}, 82\penalty0 (6):\penalty0 1063--1093, 1974.

\bibitem[Bickerton(2016)]{bickerton2016elite}
C.~Bickerton.
\newblock Europe in revolt, 12 2016.
\newblock URL \url{https://www.prospectmagazine.co.uk/essays/43719/europe-in-revolt}.

\bibitem[Blume(1993)]{blume1993statistical}
L.~E. Blume.
\newblock The statistical mechanics of strategic interaction.
\newblock \emph{Games and economic behavior}, 5\penalty0 (3):\penalty0 387--424, 1993.

\bibitem[Bollob{\'a}s and Riordan(2015)]{bollobas2015old}
B.~Bollob{\'a}s and O.~Riordan.
\newblock An old approach to the giant component problem.
\newblock \emph{Journal of Combinatorial Theory, Series B}, 113:\penalty0 236--260, 2015.

\bibitem[Bourdieu(1986)]{bourdieu1986forms}
P.~Bourdieu.
\newblock The forms of capital.
\newblock In J.~G. Richardson, editor, \emph{Handbook of Theory and Research for the Sociology of Education}, pages 241--258. Greenwood Press, New York, NY, 1986.

\bibitem[Bowles and Gintis(2002)]{bowles2002social}
S.~Bowles and H.~Gintis.
\newblock Social capital and community governance.
\newblock \emph{The economic journal}, 112\penalty0 (483):\penalty0 F419--F436, 2002.

\bibitem[Bramoull{\'e} and Kranton(2007)]{bramoulle2007public}
Y.~Bramoull{\'e} and R.~Kranton.
\newblock Public goods in networks.
\newblock \emph{Journal of Economic theory}, 135\penalty0 (1):\penalty0 478--494, 2007.

\bibitem[Britton et~al.(2021)Britton, van~der Erve, Waltmann, and Xu]{DFE_2021}
J.~Britton, L.~van~der Erve, B.~Waltmann, and X.~Xu.
\newblock London calling? higher education, geographical mobility and early-career earnings.
\newblock Technical report, Department for Education and Institute for Fiscal Studies, 2021.

\bibitem[Buera et~al.(2022)Buera, Kaboski, Rogerson, and Vizcaino]{buera2022skill}
F.~J. Buera, J.~P. Kaboski, R.~Rogerson, and J.~I. Vizcaino.
\newblock Skill-biased structural change.
\newblock \emph{The Review of Economic Studies}, 89\penalty0 (2):\penalty0 592--625, 2022.

\bibitem[Buldyrev et~al.(2010)Buldyrev, Parshani, Paul, Stanley, and Havlin]{buldyrev2010catastrophic}
S.~V. Buldyrev, R.~Parshani, G.~Paul, H.~E. Stanley, and S.~Havlin.
\newblock Catastrophic cascade of failures in interdependent networks.
\newblock \emph{Nature}, 464\penalty0 (7291):\penalty0 1025--1028, 2010.

\bibitem[Bureau(2023)]{US_census_tract_2023}
U.~C. Bureau.
\newblock United states census bureau glossary, 2023.
\newblock \url{https://www.census.gov/programs-surveys/geography/about/glossary.html} [Accessed: 3 October 2023].

\bibitem[Cabrales et~al.(2011)Cabrales, Calv{\'o}-Armengol, and Zenou]{cabrales2011social}
A.~Cabrales, A.~Calv{\'o}-Armengol, and Y.~Zenou.
\newblock Social interactions and spillovers.
\newblock \emph{Games and Economic Behavior}, 72\penalty0 (2):\penalty0 339--360, 2011.

\bibitem[Campbell(2013)]{campbell2013word}
A.~Campbell.
\newblock Word-of-mouth communication and percolation in social networks.
\newblock \emph{American Economic Review}, 103\penalty0 (6):\penalty0 2466--2498, 2013.

\bibitem[Campbell et~al.(2024{\natexlab{a}})Campbell, Thornton, and Zenou]{campbell2024strategic}
A.~Campbell, D.~Thornton, and Y.~Zenou.
\newblock Strategic diffusion: Public goods vs. public bads.
\newblock 2024{\natexlab{a}}.

\bibitem[Campbell et~al.(2024{\natexlab{b}})Campbell, Ushchev, and Zenou]{campbell2024network}
A.~Campbell, P.~Ushchev, and Y.~Zenou.
\newblock The network origins of entry.
\newblock \emph{Journal of Political Economy}, 132\penalty0 (11):\penalty0 3867--3916, 2024{\natexlab{b}}.

\bibitem[Canen et~al.(2021)Canen, Jackson, and Trebbi]{canen2021endogenous}
N.~Canen, M.~O. Jackson, and F.~Trebbi.
\newblock Endogenous networks and legislative activity.
\newblock \emph{Available at SSRN 2823338}, 2021.

\bibitem[Case and Deaton(2020)]{case2020deaths}
A.~Case and A.~Deaton.
\newblock \emph{Deaths of Despair and the Future of Capitalism}.
\newblock Princeton University Press, 2020.

\bibitem[Chay and Munshi(2012)]{chay2012black}
K.~Chay and K.~Munshi.
\newblock Black networks after emancipation: evidence from reconstruction and the great migration.
\newblock \emph{Unpublished working paper}, 2012.

\bibitem[Coleman(1988)]{coleman1988social}
J.~S. Coleman.
\newblock Social capital in the creation of human capital.
\newblock \emph{American journal of sociology}, 94:\penalty0 S95--S120, 1988.

\bibitem[Commission(2024)]{generosity2024}
T.~G. Commission.
\newblock Everyday actions, extraordinary potential: the power of giving and volunteering.
\newblock Technical report, The Giving Institute, 2024.

\bibitem[Costa and Kahn(2001)]{costa2001understanding}
D.~Costa and M.~E. Kahn.
\newblock Understanding the decline in social capital, 1952-1998, 2001.

\bibitem[Currarini et~al.(2009)Currarini, Jackson, and Pin]{currarini2009economic}
S.~Currarini, M.~O. Jackson, and P.~Pin.
\newblock An economic model of friendship: Homophily, minorities, and segregation.
\newblock \emph{Econometrica}, 77\penalty0 (4):\penalty0 1003--1045, 2009.

\bibitem[Dasaratha(2023)]{dasaratha2023innovation}
K.~Dasaratha.
\newblock Innovation and strategic network formation.
\newblock \emph{The Review of Economic Studies}, 90\penalty0 (1):\penalty0 229--260, 2023.

\bibitem[Dasgupta and Serageldin(2000)]{dasgupta2000social}
P.~Dasgupta and I.~Serageldin.
\newblock \emph{Social capital: a multifaceted perspective}.
\newblock World Bank Publications, 2000.

\bibitem[Department~for Levelling~Up and Communities(2022)]{levellingup2022}
H.~Department~for Levelling~Up and Communities.
\newblock Levelling up the united kingdom.
\newblock Technical report, HMG, 2022.
\newblock Presented to Parliament by the Secretary of State for Levelling Up, Housing and Communities by Command of Her Majesty 2 February 2022.

\bibitem[Desmond(2012)]{desmond2012disposable}
M.~Desmond.
\newblock Disposable ties and the urban poor.
\newblock \emph{American Journal of Sociology}, 117\penalty0 (5):\penalty0 1295--1335, 2012.

\bibitem[Desmond(2016)]{desmond2016evicted}
M.~Desmond.
\newblock \emph{Evicted: Poverty and Profit in the American City}.
\newblock Crown Publishing Group, 2016.

\bibitem[Dhara et~al.(2017)Dhara, van~der Hofstad, van Leeuwaarden, and Sen]{dhara2017critical}
S.~Dhara, R.~van~der Hofstad, J.~S. van Leeuwaarden, and S.~Sen.
\newblock Critical window for the configuration model: finite third moment degrees.
\newblock \emph{Electron. J. Probab}, 22\penalty0 (16):\penalty0 1--33, 2017.

\bibitem[Diekmann et~al.(2014)Diekmann, Jann, Przepiorka, and Wehrli]{diekmann2014reputation}
A.~Diekmann, B.~Jann, W.~Przepiorka, and S.~Wehrli.
\newblock Reputation formation and the evolution of cooperation in anonymous online markets.
\newblock \emph{American sociological review}, 79\penalty0 (1):\penalty0 65--85, 2014.

\bibitem[Dietz and Grimm~Jr(2019)]{dietz2019less}
N.~Dietz and R.~T. Grimm~Jr.
\newblock A less charitable nation: The decline of volunteering and giving in the united states, 2019.

\bibitem[Elliott and Golub(2019)]{elliott2019network}
M.~Elliott and B.~Golub.
\newblock A network approach to public goods.
\newblock \emph{Journal of Political Economy}, 127\penalty0 (2):\penalty0 730–776, 2019.

\bibitem[Elliott et~al.(2022)Elliott, Golub, and Leduc]{elliott2022supply}
M.~Elliott, B.~Golub, and M.~V. Leduc.
\newblock Supply network formation and fragility.
\newblock \emph{American Economic Review}, 112\penalty0 (8):\penalty0 2701--47, 2022.

\bibitem[Elliott et~al.(2023)Elliott, Golub, and Leduc]{elliott2023corporate}
M.~Elliott, B.~Golub, and M.~V. Leduc.
\newblock Corporate culture and organizational fragility.
\newblock \emph{arXiv preprint arXiv:2301.08907}, 2023.

\bibitem[Ellison(1994)]{ellison1994cooperation}
G.~Ellison.
\newblock Cooperation in the prisoner's dilemma with anonymous random matching.
\newblock \emph{The Review of Economic Studies}, 61\penalty0 (3):\penalty0 567--588, 1994.

\bibitem[{European Commision}(2020)]{EC2020skills}
{European Commision}.
\newblock European skills agenda for sustainable competitiveness, social fairness and resilience.
\newblock Technical report, European Commision, 2020.

\bibitem[Farboodi et~al.(2023)Farboodi, Jarosch, and Shimer]{farboodi2023emergence}
M.~Farboodi, G.~Jarosch, and R.~Shimer.
\newblock The emergence of market structure.
\newblock \emph{The Review of Economic Studies}, 90\penalty0 (1):\penalty0 261--292, 2023.

\bibitem[Fehr and G{\"a}chter(2002)]{fehr2002altruistic}
E.~Fehr and S.~G{\"a}chter.
\newblock Altruistic punishment in humans.
\newblock \emph{Nature}, 415\penalty0 (6868):\penalty0 137--140, 2002.

\bibitem[Fehr and Gintis(2007)]{fehr2007human}
E.~Fehr and H.~Gintis.
\newblock Human motivation and social cooperation: Experimental and analytical foundations.
\newblock \emph{Annu. Rev. Sociol.}, 33:\penalty0 43--64, 2007.

\bibitem[Feinberg et~al.(2012)Feinberg, Willer, Stellar, and Keltner]{feinberg2012virtues}
M.~Feinberg, R.~Willer, J.~Stellar, and D.~Keltner.
\newblock The virtues of gossip: reputational information sharing as prosocial behavior.
\newblock \emph{Journal of personality and social psychology}, 102\penalty0 (5):\penalty0 1015, 2012.

\bibitem[Fountoulakis et~al.(2022)Fountoulakis, Joos, and Perarnau]{fountoulakis2022percolation}
N.~Fountoulakis, F.~Joos, and G.~Perarnau.
\newblock Percolation on random graphs with a fixed degree sequence.
\newblock \emph{SIAM Journal on Discrete Mathematics}, 36\penalty0 (1):\penalty0 1--46, 2022.

\bibitem[Funk(2010)]{funk2010social}
P.~Funk.
\newblock Social incentives and voter turnout: evidence from the swiss mail ballot system.
\newblock \emph{Journal of the European economic association}, 8\penalty0 (5):\penalty0 1077--1103, 2010.

\bibitem[Galston(2018)]{galston2018populist}
W.~A. Galston.
\newblock The populist challenge to liberal democracy.
\newblock \emph{Journal of Democracy}, 29\penalty0 (2):\penalty0 5--19, 2018.

\bibitem[Glaeser et~al.(2002)Glaeser, Laibson, and Sacerdote]{glaeser2002economic}
E.~L. Glaeser, D.~Laibson, and B.~Sacerdote.
\newblock An economic approach to social capital.
\newblock \emph{The economic journal}, 112\penalty0 (483):\penalty0 F437--F458, 2002.

\bibitem[Granovetter(1978)]{granovetter1978threshold}
M.~Granovetter.
\newblock Threshold models of collective behavior.
\newblock \emph{American journal of sociology}, 83\penalty0 (6):\penalty0 1420--1443, 1978.

\bibitem[Guiso et~al.(2006)Guiso, Sapienza, and Zingales]{guiso2006does}
L.~Guiso, P.~Sapienza, and L.~Zingales.
\newblock Does culture affect economic outcomes?
\newblock \emph{Journal of Economic perspectives}, 20\penalty0 (2):\penalty0 23--48, 2006.

\bibitem[Guiso et~al.(2011)Guiso, Sapienza, and Zingales]{guiso2011civic}
L.~Guiso, P.~Sapienza, and L.~Zingales.
\newblock Civic capital as the missing link.
\newblock \emph{Handbook of social economics}, 1:\penalty0 417--480, 2011.

\bibitem[Hargadon(2003)]{hargadon2003breakthroughs}
A.~Hargadon.
\newblock \emph{How breakthroughs happen: The surprising truth about how companies innovate}.
\newblock Harvard Business Press, 2003.

\bibitem[Hatami and Molloy(2012)]{hatami2012scaling}
H.~Hatami and M.~Molloy.
\newblock The scaling window for a random graph with a given degree sequence.
\newblock \emph{Random Structures \& Algorithms}, 41\penalty0 (1):\penalty0 99--123, 2012.

\bibitem[Jackson(2008)]{jackson2008}
M.~O. Jackson.
\newblock \emph{Social and economic networks}.
\newblock Princeton University Press, 2008.

\bibitem[Jackson(2020)]{jackson2020typology}
M.~O. Jackson.
\newblock A typology of social capital and associated network measures.
\newblock \emph{Social choice and welfare}, 54\penalty0 (2-3):\penalty0 311--336, 2020.

\bibitem[Jackson et~al.(2012)Jackson, Rodriguez-Barraquer, and Tan]{jackson2012social}
M.~O. Jackson, T.~Rodriguez-Barraquer, and X.~Tan.
\newblock Social capital and social quilts: Network patterns of favor exchange.
\newblock \emph{American Economic Review}, 102\penalty0 (5):\penalty0 1857--97, 2012.

\bibitem[Jackson et~al.(2023)Jackson, Nei, Snowberg, and Yariv]{jackson2023dynamics}
M.~O. Jackson, S.~M. Nei, E.~Snowberg, and L.~Yariv.
\newblock The dynamics of networks and homophily.
\newblock Technical report, National Bureau of Economic Research, 2023.

\bibitem[Janson(2009)]{janson2009percolation}
S.~Janson.
\newblock On percolation in random graphs with given vertex degrees.
\newblock \emph{Electronic Journal of Probability}, 14\penalty0 (5):\penalty0 87--118, 2009.

\bibitem[Janson and Luczak(2009)]{janson2009new}
S.~Janson and M.~J. Luczak.
\newblock A new approach to the giant component problem.
\newblock \emph{Random Structures \& Algorithms}, 34\penalty0 (2):\penalty0 197--216, 2009.

\bibitem[Johnson and Acemoglu(2023)]{johnson2023power}
S.~Johnson and D.~Acemoglu.
\newblock \emph{Power and Progress: Our Thousand-Year Struggle Over Technology and Prosperity}.
\newblock Hachette UK, 2023.

\bibitem[Kandori(1992)]{kandori1992social}
M.~Kandori.
\newblock Social norms and community enforcement.
\newblock \emph{The Review of Economic Studies}, 59\penalty0 (1):\penalty0 63--80, 1992.

\bibitem[Langtry et~al.(2024)Langtry, Taylor, and Zhang]{langtry2024network}
A.~Langtry, S.~Taylor, and Y.~Zhang.
\newblock Network threshold games.
\newblock \emph{arXiv preprint arXiv:2406.04540}, 2024.

\bibitem[Lippert and Spagnolo(2011)]{lippert2011networks}
S.~Lippert and G.~Spagnolo.
\newblock Networks of relations and word-of-mouth communication.
\newblock \emph{Games and Economic Behavior}, 72\penalty0 (1):\penalty0 202--217, 2011.

\bibitem[List(2006)]{list2006behavioralist}
J.~A. List.
\newblock The behavioralist meets the market: Measuring social preferences and reputation effects in actual transactions.
\newblock \emph{Journal of political Economy}, 114\penalty0 (1):\penalty0 1--37, 2006.

\bibitem[Mas and Moretti(2009)]{mas2009peers}
A.~Mas and E.~Moretti.
\newblock Peers at work.
\newblock \emph{American Economic Review}, 99\penalty0 (1):\penalty0 112--145, 2009.

\bibitem[Matouschek et~al.(2025)Matouschek, Powell, and Reich]{matouschek2025organizing}
N.~Matouschek, M.~Powell, and B.~Reich.
\newblock Organizing modular production.
\newblock \emph{Journal of Political Economy}, 133\penalty0 (3), 2025.

\bibitem[May(2017)]{may2017}
T.~May.
\newblock The shared society, 1 2017.
\newblock URL \url{https://www.gov.uk/government/speeches/the-shared-society-prime-ministers-speech-at-the-charity-commission-annual-meeting}.
\newblock Prime Minister's speech at the Charity Commission annual meeting [Accessed: 2025 02 25].

\bibitem[Melnikov et~al.(2020)Melnikov, Schmidt-Padilla, and Sviatschi]{melnikov2020gangs}
N.~Melnikov, C.~Schmidt-Padilla, and M.~M. Sviatschi.
\newblock Gangs, labor mobility and development.
\newblock Technical report, National Bureau of Economic Research, 2020.

\bibitem[Molloy and Reed(1995)]{molloy1995critical}
M.~Molloy and B.~Reed.
\newblock A critical point for random graphs with a given degree sequence.
\newblock \emph{Random structures \& algorithms}, 6\penalty0 (2-3):\penalty0 161--180, 1995.

\bibitem[Molloy and Reed(1998)]{molloy1998size}
M.~Molloy and B.~Reed.
\newblock The size of the giant component of a random graph with a given degree sequence.
\newblock \emph{Combinatorics, probability and computing}, 7\penalty0 (3):\penalty0 295--305, 1998.

\bibitem[Morris(2000)]{morris2000contagion}
S.~Morris.
\newblock Contagion.
\newblock \emph{The Review of Economic Studies}, 67\penalty0 (1):\penalty0 57--78, 2000.

\bibitem[Nava and Piccione(2014)]{nava2014efficiency}
F.~Nava and M.~Piccione.
\newblock Efficiency in repeated games with local interaction and uncertain local monitoring.
\newblock \emph{Theoretical Economics}, 9\penalty0 (1):\penalty0 279--312, 2014.

\bibitem[Newman(2009)]{newman2009random}
M.~E. Newman.
\newblock Random graphs with clustering.
\newblock \emph{Physical review letters}, 103\penalty0 (5):\penalty0 058701, 2009.

\bibitem[OECD(2019)]{OECD2019future}
OECD.
\newblock Getting skills right: Future-ready adult learning systems.
\newblock Technical report, Organisation for Economic Co-operation and Development, 2019.

\bibitem[Probst and Scharff(2019)]{PWC2019skills}
L.~Probst and C.~Scharff.
\newblock A strategist’s guide to upskilling.
\newblock Technical report, PWC, 2019.
\newblock URL \url{https://www.strategy-business.com/feature/A-strategists-guide-to-upskilling}.

\bibitem[Putnam(1995)]{putnam1995tuning}
R.~D. Putnam.
\newblock Tuning in, tuning out: The strange disappearance of social capital in america.
\newblock \emph{PS: Political science \& politics}, 28\penalty0 (4):\penalty0 664--683, 1995.

\bibitem[Putnam(2000)]{putnam2000bowling}
R.~D. Putnam.
\newblock \emph{Bowling alone: The collapse and revival of American community}.
\newblock Simon and schuster, 2000.

\bibitem[Putnam(2016)]{putnam2016our}
R.~D. Putnam.
\newblock \emph{Our kids: The American dream in crisis}.
\newblock Simon and Schuster, 2016.

\bibitem[Rosen(1965)]{rosen1965existence}
J.~B. Rosen.
\newblock Existence and uniqueness of equilibrium points for concave n-person games.
\newblock \emph{Econometrica: Journal of the Econometric Society}, pages 520--534, 1965.

\bibitem[Sadler(2020)]{sadler2020diffusion}
E.~Sadler.
\newblock Diffusion games.
\newblock \emph{American Economic Review}, 110\penalty0 (1):\penalty0 225--270, 2020.

\bibitem[Sadler(2024)]{sadler2024simple}
E.~Sadler.
\newblock Simple conditions for a unique nash equilibrium.
\newblock \emph{Available at SSRN}, 2024.

\bibitem[Simpson and Willer(2015)]{simpson2015beyond}
B.~Simpson and R.~Willer.
\newblock Beyond altruism: Sociological foundations of cooperation and prosocial behavior.
\newblock \emph{Annual Review of Sociology}, 41:\penalty0 43--63, 2015.

\bibitem[Sirangelo(2024)]{forbes2024volunteering}
J.~Sirangelo.
\newblock The decline in volunteering is driving increased division, 3 2024.
\newblock URL \url{https://www.forbes.com/councils/forbesnonprofitcouncil/2024/03/29/the-decline-in-volunteering-is-driving-increased-division/}.

\bibitem[Sprung-Keyser et~al.(2022)Sprung-Keyser, Hendren, Porter, et~al.]{sprung2022radius}
B.~Sprung-Keyser, N.~Hendren, S.~Porter, et~al.
\newblock \emph{The radius of economic opportunity: Evidence from migration and local labor markets}.
\newblock US Census Bureau, Center for Economic Studies, 2022.

\bibitem[Stolle and Hooghe(2005)]{stolle2005inaccurate}
D.~Stolle and M.~Hooghe.
\newblock Inaccurate, exceptional, one-sided or irrelevant? the debate about the alleged decline of social capital and civic engagement in western societies.
\newblock \emph{British Journal of Political Science}, 35\penalty0 (1):\penalty0 149--167, 2005.

\bibitem[Tanner et~al.(2020)Tanner, O’Shaughnessy, Krasniqi, and Blagden]{tanner2020state}
W.~Tanner, J.~O’Shaughnessy, F.~Krasniqi, and J.~Blagden.
\newblock The state of our social fabric, 2020.

\bibitem[Thornton(2019)]{thornton2019}
D.~Thornton.
\newblock \emph{The Giant Component: Random Graphs and Information Cascades}.
\newblock PhD thesis, UNSW Sydney, 2019.

\bibitem[Vance(2022)]{vance2022hillbilly}
J.~D. Vance.
\newblock \emph{Hillbilly elegy}.
\newblock William Collins, 2022.

\bibitem[Vega-Redondo(2006)]{vega2006building}
F.~Vega-Redondo.
\newblock Building up social capital in a changing world.
\newblock \emph{Journal of Economic Dynamics and Control}, 30\penalty0 (11):\penalty0 2305--2338, 2006.

\bibitem[Violante(2008)]{violante2008skill}
G.~L. Violante.
\newblock Skill-biased technical change.
\newblock \emph{The new Palgrave dictionary of economics}, 2:\penalty0 1--6, 2008.

\bibitem[Watts(2002)]{watts2002simple}
D.~J. Watts.
\newblock A simple model of global cascades on random networks.
\newblock \emph{Proceedings of the National Academy of Sciences}, 99\penalty0 (9):\penalty0 5766--5771, 2002.

\bibitem[Willer(2009)]{willer2009groups}
R.~Willer.
\newblock Groups reward individual sacrifice: The status solution to the collective action problem.
\newblock \emph{American Sociological Review}, 74\penalty0 (1):\penalty0 23--43, 2009.

\bibitem[Zhang and Zhu(2011)]{zhang2011group}
X.~Zhang and F.~Zhu.
\newblock Group size and incentives to contribute: A natural experiment at chinese wikipedia.
\newblock \emph{American Economic Review}, 101\penalty0 (4):\penalty0 1601--1615, 2011.

\end{thebibliography}
\cleardoublepage
\onehalfspacing
\appendix
\numberwithin{equation}{section}
\numberwithin{figure}{section}
\numberwithin{thm}{section}
\numberwithin{prop}{section}
\numberwithin{defn}{section}
\numberwithin{rem}{section}
\numberwithin{lem}{section}
\numberwithin{cor}{section}
\newpage

\section{Proofs}\label{sec:proofs}
\paragraph{A technical condition}
The proof to \Cref{prop:second_stage} uses a result from 
\cite{janson2009new}, which is stated subject to the follow technical condition.

\begin{cond}[\cite{janson2009new}, Condition 2.1]\label{ass:degree_dist_convergence}
for each $n$, $(d_i)_1^n = (d_i^{(n)})_1^n$ is a sequence of of non-negative integers such that $\sum_{i=1}^n d_i$ is even. Furthermore, $(\lambda_d)_{d=0}^{\infty}$ is a probability distribution independent of $n$ such that: \\
(i) \ \ $n_d / n = |\{i: d_i = d\}| / n \to \lambda_d$ as $n \to \infty$, for every $d \geq 0$; \\
(ii)  \ $\sum_d d \lambda_d \in (0,\infty)$; 
(iii) $\sum_i d_i^2 = \mathcal{O}(n)$; 
(iv) \ $\lambda_1 > 0$.
\end{cond}

Part (i) is assumed in the statement of \Cref{prop:second_stage}. Parts (ii) and (iii) are implied by the (stronger) assumption that all agents have finite degree (assumption (2) in the \emph{network formation} part of \Cref{sec:model}). Part (iv) follows directly from assumption (3) in the \emph{network formation} part of \Cref{sec:model}. The evenness of $\sum_{i=1}^n d_i$ is also assumed directly.
%
In the interest of completeness, I also state some standard definitions that are used in the proof.

\begin{defn}[configuration model]
    A \emph{configuration model}, $\mathcal{G}(\mathbf{d}^{(n)})$, draws an unweighted and undirected graph, $G$, uniformly at random from those with the degree sequence $\mathbf{d}^{(n)}$.\footnote{Formally, this is a configuration model specified for a simple graph -- one with no self-loops or multiple edges. It is possible to allow self-loops and/or multiple edges. The uniform sampling remains unchanged. Notice that this process matches the specified degree distribution exactly. Formally, this is called a micro-canonical configuration model. The other type, the canonical model, preserves only the expected degree sequence. See \cite{thornton2019} for excellent discussion.}    
\end{defn}

\begin{defn}[giant component]
    A component, $\mathcal{C}$, of a network $G$, is a \emph{giant component} if $v(\mathcal{C})/n \overset{p}{\to} k$ as $n \to \infty$ for some $k>0$, where $v(\mathcal{C})$ is the number of agents in $\mathcal{C}$.
\end{defn}

\begin{defn}[w.h.p]
An event occurs \emph{with high probability} (w.h.p.) if the probability the event occurs goes to $1$ as $n$ goes to $\infty$. Formally: $X$ occurs w.h.p. if and only if $\mathbb{P}(X) \to 1$ as $n \to \infty$.    
\end{defn}

\subsection*{Proof of Proposition 1} 
It follows immediately from agents' preferences (equation \ref{eq:prefs}) that $x_i^* = 1$ if and only if $\mathbb{P}_i(\text{reward}) \cdot R \geq C_i$. The remainder of the proof amounts to characterising $p_i(\text{reward})$. First, I show that if $G(i)$ has a \emph{giant component} then $i$ gets a reward for her public good provision \emph{w.h.p.}. Second, if $G(i)$ does \emph{not} have a giant component then $i$ does not a reward for her public good provision \emph{w.h.p.}. Third, I apply an existing result that shows that the giant component exists \emph{w.h.p.} if $\sum_d d (d - 2) \lambda_d > 0$, and does not exist \emph{w.h.p.} otherwise.

\paragraph{Step 1: Giant component $\implies$ $p_i(\text{reward}) \to 1$ as $n \to \infty$.}
Agent $i$ receives the reward if and only if \emph{at least one} observer and \emph{at least one} rewarder are in the same connected component. A sufficient condition is therefore that at least one and rewarder are both in the giant component.

Let $C_1(G)$ be the set of agents in the \emph{largest} component of $G$. If a giant component exists, it is clear that the largest component must be giant.\footnote{It is also well-known that there can only be one giant component (see, for example, \cite{janson2009new}.} 
Therefore we have
\begin{align}
    p_i(\text{reward} | \text{GC}) &\geq [1 - \mathbb{P}(\mathcal{N}_i^o \cap C_1(G) = \emptyset)] \times [1 - \mathbb{P}(\mathcal{N}_i^r \cap C_1(G) = \emptyset)].
\end{align}

Recall that the networks $\mathcal{N}_i^o$ and $\mathcal{N}_i^r$ are random sets of agents. So the probability no observers are in the giant component is just the product of the probabilities that each observer is not in the giant component. That is, $\mathbb{P}(\mathcal{N}_i^r \cap C_1(G) = \emptyset) = \mathbb{P}(j \notin C_1(G))^{| \mathcal{N}_i^r |}$. And the probability that one observer is \emph{not} in the largest component is just equal to the fraction of all agents not in the giant component. That is, $\mathbb{P}(j \notin C_1(G)) = 1 - \frac{1}{n} C_1(G)$. 

By definition, the presence of a giant component in the network $\mathcal{G}$ means that $\frac{1}{n} C_1(\mathcal{G}) \overset{p}{\to} k$ as $n \to \infty$, for some $k > 0$ (see \cite{janson2009new}). Now recall that by assumption $| \mathcal{N}_i^o | \to \infty$ as $n \to \infty$. This is because I assumed that $\mathcal{O}(n^{\alpha})$), for some $\alpha \in (0, \frac{1}{6})$. Therefore $\mathbb{P}(\mathcal{N}_i^o \cap C_1(\mathcal{G}) = \emptyset) \to (1 - \alpha)^{| \mathcal{N}_i^o |} \to 0$ as $n \to \infty$. Identical reasoning holds for $\mathbb{P}(\mathcal{N}_i^r \cap C_1(G) = \emptyset)$. Together, these yield: $p_i(\text{reward} | \text{GC}) \to 1$ as $n \to \infty$: that is, $i$ gets the reward \emph{w.h.p.}.

\paragraph{Step 2: \emph{No} giant component $\implies$ $p_i(\text{reward}) \to 0$ as $n \to \infty$.} The best case scenario in terms of $i$ getting a reward is when all of her observers are in different component of the gossip network \emph{and} all of those components are just as large as the network's largest component. In this case, a fraction $\min \{1, | \mathcal{N}_i^o | \cdot |C_1| \cdot n^{-1}\}$ of all agents learn $i$'s action.\footnote{We need the minimum function because this naive way of constructing a best case scenario does not guarantee a fraction below $1$.} 

Since rewarders are randomly selected, the probability that a given rewarder \emph{does not} learn $i$'s action is at least $(1 - \min \{1, | \mathcal{N}_i^o | \cdot |C_1| \cdot n^{-1}\})$. Then the probability that \emph{all} rewarders \emph{do not} learn $i$'s action is then at least $(1 - \min \{1, | \mathcal{N}_i^o | \cdot |C_1| \cdot n^{-1}\})^{|\mathcal{N}_i^r|}$. That is,
\begin{align}
    p_i(\text{no reward} | \text{No GC}) \geq (1 - \min \{1, | \mathcal{N}_i^o | \cdot |C_1| \cdot n^{-1}\})^{|\mathcal{N}_i^r|}.
\end{align}

We can write $| \mathcal{N}_i^o | = \beta^o n^{\sfrac{1}{6}}$ and $|\mathcal{N}_i^r| = \beta^r n^{\sfrac{1}{6}}$ for $\beta^o, \beta^r \to 0$ as $n \to \infty$. This is because we assumed that $| \mathcal{N}_i^o |$ and $| \mathcal{N}_i^r |$ are $\mathcal{O}(n^{\alpha})$ for some $\alpha \in (0, \frac{1}{6})$. Then a known result about random networks is that when a giant component does not exist, the largest component is \emph{at most} on the order of $n^{\sfrac{2}{3}}$ \citep{hatami2012scaling,dhara2017critical}. So $|C_1| = \zeta_n n^{\sfrac{2}{3}}$, for $\{\zeta\}$, $\zeta_n \to K$ as $n \to \infty$ where $K$ is some non-negative real number.\footnote{In fact, components of this size only occur in what is known as the `critical phase', where a giant component is on the cusp of appearing. In the `subcritical phase'---the conventional state for random networks without a giant component---the largest component is actually on the order of $\ln(n)$ \citep[esp. Lemma 3]{molloy1995critical}. So this bound is extremely loose in all but the `critical phase' case.}

Using these yields $p_i(\text{no reward}| \text{No GC}) \geq (1 - \min \{1, \beta^o n^{\sfrac{1}{6}} \zeta_n n^{\sfrac{2}{3}} n^{-1}\})^{\beta^r n^{\sfrac{1}{6}}}$. And notice that for sufficiently large $n$, we must have $\beta^o n^{\sfrac{1}{6}} \zeta_n n^{\sfrac{2}{3}} n^{-1} < 1$. So we can safely ignore the min function. Then $p_i(\text{no reward}| \text{No GC}) \geq (1 - \beta^o \zeta_n n^{\sfrac{-1}{6}})^{\beta^r n^{\sfrac{1}{6}}}$. 
And by the definition of the exponential function, we have
\begin{align}
    (1 - \beta^o \zeta_n n^{\sfrac{-1}{6}})^{\beta^r n^{\sfrac{1}{6}}} \to e^{-\beta^o \beta^r \zeta_n} \text{  as 
 } n \to\infty.
\end{align}
Therefore $\lim_{n \to \infty} p_i(\text{no reward}| \text{No GC}) \geq \lim_{n \to \infty} e^{-\beta^o \beta^r \zeta_n}$. But we also know that $\beta^o \beta^r \zeta_n \to 0$ as $n \to \infty$ (and $e^a \to 1$ as $a \to 0$). So $p_i(\text{no reward}| \text{No GC}) \to 1$ as $n \to \infty$ (as a probabilities cannot go above $1$). Which finally yields $p_i(\text{reward}| \text{No GC}) \to 0$ as $n \to \infty$: that is, $i$ does not get the reward \emph{w.h.p.}.

\paragraph{Step 3: Emergence of the giant component.} \citet[Theorem 2.3]{janson2009new} prove that if Condition 2.1 is met, then a giant component exists \emph{w.h.p.} if $\sum_d d (d - 2) \lambda_d > 0$, and does not exist \emph{w.h.p.} otherwise. Recall that the $\lambda_d$'s are defined as $\plim_{n \to \infty} \frac{1}{n} | \{ j : d_j = d \}| = \lambda_d$. In other words, we have $\mathbb{P}(\text{GC}) \to 1$ as $n \to \infty$ if $\sum_d d (d - 2) \lambda_d > 0$, and $\mathbb{P}(\text{GC}) \to 0$ as $n \to \infty$ otherwise. It is clear that excluding a single agent $i$ from the network does not change the anything. Formally, if  $\plim_{n \to \infty} \frac{1}{n} | \{ j : d_j = d \}| = \lambda_d$, then  $\plim_{n \to \infty} \frac{1}{n} | \{ j : d_j = d, j \neq i \}| = \lambda_d$. So the result applies equally to the configuration model $\mathcal{G}(\mathbf{d_{-i}})$ as it does to $\mathcal{G}(\mathbf{d})$.

\paragraph{Putting things together.} By definition 
\begin{align}
  p_i(\text{reward}) = \underbrace{p_i(\text{reward}| \text{GC})}_{\to 1 \text{ as } n \to \infty} \times \mathbb{P}(\text{GC})
  + \underbrace{p_i(\text{reward}| \text{No GC})}_{\to 0 \text{ as } n \to \infty} \times \mathbb{P}(\text{No GC})  
\end{align}
Therefore we have $p_i(\text{reward}) \to 1$ as $n \to \infty$ if $\sum_d d (d - 2) \lambda_d > 0$, and $p_i(\text{reward}) \to 0$ as $n \to \infty$ otherwise.

\paragraph{Summing up.} Firstly, in the case where $C_i = R$, the equilibrium action is $x_i^* = 0$ because there is no $n$ sufficiently large such that $p_i(\text{reward}) = 1$ (it can only tend towards $1$ as $n \to \infty$). Similarly, when $C_i = 0$, equilibrium action is $x_i^* = 0$ because there is no $n$ sufficiently large such that $p_i(\text{reward}) = 0$ (it can only tend towards $0$ as $n \to \infty$).  

The fact that cost parameters, $C_i$, are drawn from a Probability Mass Function with finite support is now helpful. It means that there exists $\mu \in (\frac{1}{2}, 1)$ such that all agents with $C_i \in (0,R)$ in fact have $C_i \in [(1 - \mu), \mu R]$. And then for any fixed $\mu \in (\frac{1}{2},1)$, we can have $n_0$ sufficiently large that $p_i(\text{reward}) \geq \mu$ when $\sum_d d(d-2) \lambda_d > 0$ and $p_i(\text{reward}) < (1 - \mu)$ when $\sum_d d(d-2) \lambda_d \leq 0$.\footnote{If we had cost parameters drawn from some continuous distribution, it would only be possible to characterise the behaviour of agents with costs $C_i \in [(1 - \mu), \mu R]$ for some fixed $\mu \in (\frac{1}{2}, 1)$. While we could choose $\mu$ as close to 1 as we like, it is not possible to characterise the behaviour of all agents. Picking some $\mu$ closer to one would require a larger $n_0$ -- i.e., more agents. But if $F(C)$ is continuous, this creates the possibility that taking draws of the cost parameter (i.e. having more agents) creates moves the `closest positive cost' (i.e. $\min_{i \in N} \{C_i : C_i > 0\}$ closer to zero. } 
Finally notice that for all $i$ with $C_i \in [(1 - \mu)R, \mu R]$, $p_i(\text{reward}) \geq \mu$ is sufficient for $x_i^* = 1$, and similarly, $p_i(\text{reward}) < (1 - \mu)$ is sufficient for $x_i^* = 0$. \hfill $\qed$

\subsection*{A Lemma on convergent behaviour} 
Before proving \Cref{prop:first_stage}, I first show that if agents' choices are `convergent' in the sense of \Cref{defn:convergent_NE} then the condition required for characterising second stage behaviour is satisfied. This is an important step because we can only characterise first stage behaviour if we know what happens in the second stage.

\begin{lem}\label{lem:technical_1}
    Suppose $\{(t_i)_{i=1}^n \}_{n \in \mathbb{N}}$ is such that $\Lambda^{(n)}(t) \overset{d}{\to} \Lambda(t)$ as $n \to \infty$. Then \Cref{ass:degree_dist_convergence} holds.
\end{lem}

\noindent \emph{Proof.} \textbf{Step 1.} Each agent's degree is a random draw from the distribution $p(d|t_i)$. From this, for each $i$ and each $d$, define a random variable $Z_{d,i}(t_i)$, with $Z_{d,i}(t_i) = 1$ if $d_i = d$, and $Z_{d,i}(t_i) = 0$ otherwise. It is clear that $Z_{d,i}(t_i)$ follows a Bernoulli distribution with $\mathbb{E}[Z_{d,i}(t_i)] = p(d|t_i)$. Then the expected fraction of agents with degree $d$ is equal to $\frac{1}{n} \sum_{i=1}^n \mathbb{E}[Z_{d,i}(t_i)] = \frac{1}{n} \sum_{i=1}^n p(d|t_i)$. 

\paragraph{Step 2: Kolmogorov's Strong LLN.} For a given $d$, realisations of $Z_{d,i}(t_i)$ are independent draws from \emph{different (i.e. not identically distributed)} Bernoulli distributions. Nevertheless, by Kolmogorov’s strong law of large numbers, we have,
\begin{align}
    \frac{1}{n} \sum_{i=1}^n Z_{d,i}(t_i) \overset{a.s.}{\to} \frac{1}{n} \sum_{i=1}^n \mathbb{E}[Z_{d,i}(t_i)]
\end{align}
This is possible because the Kolmogorov Criterion is met: namely $\sum_{i=1}^\infty \frac{\sigma_{d,i}^2}{i^2} < \infty$, where $\sigma_{d,i}^2$ is the variance of the random variable $Z_{d,i}(t_i)$. To see this, notice that all random variables here follow a Bernoulli distribution. Therefore $\sigma_{d,i}^2 = \mathbb{E}[Z_{d,i}(t_i)] (1 - \mathbb{E}[Z_{d,i}(t_i)]) \leq 0.25$. So we have 
\begin{align}
    \sum_{i=1}^\infty \frac{\sigma_{d,i}^2}{i^2} \leq 0.25 \sum_{i=1}^\infty \frac{1}{i^2} < \infty
\end{align}
The final summation is called the `Basel problem'.\footnote{It is a special case of the Riemann-Zeta function, and was first proved by Euler.} Many proofs exist.
Therefore the fraction of agents with degree $d$ converges to the \emph{expected} fraction.

\paragraph{Step 3: Considering convergent $t$.} We are only considering sequences of choices $\{(t_i)_{i=1}^n \}_{n \in \mathbb{N}}$ such that $\Lambda^{(n)}(t) \overset{d}{\to} \Lambda(t)$ as $n \to \infty$. Given this, $\frac{1}{n} \sum_{i=1}^n p(d|t_i) \to \int_0^1 p(d | t) \partial \Lambda(t)$. This is by a standard law of large numbers. We can view $p(d|t_i)$ as a function of $t_i$, with each $t_i$ an i.i.d. draw from $\Lambda(t)$. The right hand side is the usual expected value. And let $\lambda_d := \int_0^1 p(d | t) \partial \Lambda(t)$.

\paragraph{Step 4: Summing up.} From step 1, we have that the actual fraction of agents with degree $d$ converges to its expectation. And from step 2, we have that the expectation converges to $\lambda_d$. So
\begin{align}
    \frac{1}{n} |\{ i : d_i = d\}| \to \lambda_d \text{ for all } d.
\end{align}

So \Cref{ass:degree_dist_convergence}(i) is satisfied. Parts (ii)-(iv) follow from other assumptions (see discussion immediately following \Cref{ass:degree_dist_convergence}). \hfill $\qed$

\subsection*{Proof of Proposition 2}
First, I show that if \Cref{ass:degree_dist_convergence} holds, then for large enough $n$ (i.e. $n> n_0$) an agent's first stage choice of $t_i, t_i^{out}$ has no impact on the second stage payoffs. Second, I use this to characterise equilibrium behaviour in the first stage for $n> n_0$ when \Cref{ass:degree_dist_convergence} holds. Third, I show existing by noticing that this characterisation of equilibrium behaviour in fact satisfies \Cref{ass:degree_dist_convergence}.


\paragraph{Step 1A: If \Cref{ass:degree_dist_convergence} holds, $i$ has no impact on $\lim_{n \to \infty} p_j(\text{reward})$.} By assumption an agent's degree can only vary between $0$ and $D$. By \Cref{ass:degree_dist_convergence}, we have $\frac{1}{n} |\{i : d_i = d\}| \overset{p}{\to} \lambda_d$ for all $d$. Changing the degree of a single agent $\ell$ cannot affect this. Formally; if $\frac{1}{n} |\{i : d_i = d\}| \overset{p}{\to} \lambda_d$ for all $d$ then we also have $\frac{1}{n} \left( |\{i : d_i = d\}| \pm 1 \right) \overset{p}{\to} \lambda_d$ for all $d$. This means that the results from Proposition 1 do not change when we vary the degree of exactly 1 agent $\ell$.\footnote{The case for $j=i$ is in fact trivial: $i$ does not form part of her own network, $G(i)$, and so cannot affect $p_i(\text{reward})$.}
Now let us unpack this more carefully. Consider exactly what happens when $\ell$ changes her behaviour. 

\emph{Case A: a giant component exists} (which it does with high probability when $\sum_d d(d-2) \lambda_d > 0$). Then there is one component containing at least $k n$ nodes for some $k > 0$ (the giant component), and all other components contain at most $\mathcal{O}(\ln n)$ nodes.\footnote{The fact that there is only one giant component, and that all non-giant components are of size at most $\mathcal{O}(\ln n)$ is a standard result. See for example \cite{molloy1995critical, molloy1998size}.} 
Increasing $\ell$'s degree can only create larger components -- which trivially cannot destroy the giant component. 
Decreasing $\ell$'s degree could break up components. If $\ell$ is not in the giant component, this trivially has no effect on the existence of the giant component. If $i$ \emph{is} in the giant component, decreasing her degree can at worst break up the giant component into $D$ different pieces (because $i$'s degree cannot decrease by more than $D$ by construction). But one of these pieces must be of size at least $\frac{1}{D} k n$. And this component is still on the order of $n$, and so is still a giant component. 

\emph{Case B: a giant component does not exist} (which it does with high probability when $\sum_d d(d-2) \lambda_d \leq 0$). All components are at most on the order of $n^{\sfrac{2}{3}}$. Decreasing $i$'s degree can only break up components, and reduce $p_j(\text{reward})$ -- which are already tending to 0. 
Increasing $\ell$'s degree can at most connect $D$ different components, all of size $\mathcal{O}(n^{\sfrac{2}{3}})$. But this enlarged component is also $\mathcal{O}(n^{\frac{2}{3}})$.\footnote{Recall that by definition a component, $\mathcal{C}$, is $\mathcal{O}(n^{\sfrac{2}{3}})$ if for all $k>0$, there exists an $n_0$ such that for all $n\geq n_0$, $|\mathcal{C}| \leq k n^{\sfrac{2}{3}}$.} 

Finally, we know from \Cref{prop:second_stage} that $\lim_{n \to \infty} p_j(\text{reward})$ only depends on whether or not a giant component exists. Therefore, as $i$ can have no impact on the existence of a giant component, $i$ can have no impact on $\lim_{n \to \infty} p_j(\text{reward})$.

\paragraph{Step 1B: If \Cref{ass:degree_dist_convergence} holds, then for $n_0 > n$, $i$ has no impact on the second stage.} 
As $i$'s degree does not affect $\lim_{n \to \infty} p_j(\text{reward})$, for any $\mu \in (\frac{1}{2}, 1)$, there must exist an $n_0$ such that for all $n>n_0$, $p_j(\text{reward}) \geq \mu$ if $\sum_d d(d-2) \lambda_d > 0$ and $p_j(\text{reward}) \leq 1-\mu$ if $\sum_d d(d-2) \lambda_d \leq 0$.
And we also know from the proof to \Cref{prop:second_stage} that there exists a $\mu \in (\frac{1}{2}, 1)$ such that all agents with $C_i \in (0,R)$ in fact have $C_i \in [(1 - \mu), \mu R]$. Taking the $n_0$ associated with this value of $\mu$, for all $n>n_0$, all agents' second stage behaviour depends only on the the existence or otherwise of a giant component (i.e. on the value of $\sum_d d(d-2)\lambda_d$). In sum, for sufficient large $n$ (i.e. $n> n_0$), $i$'s second stage payoffs
\begin{align}
    v_i^* = \max_{x_{i} \in \{0,1\}} \{ p_{i}(\text{reward}) R x_i - C_{i} x_{i} \} + B \left(\frac{1}{n-1} \sum_{j \neq \ell} x_j^* \right)
\end{align}

do not depend on her own degree.

\paragraph{Step 1C: mapping changes in $t_i$ into changes in $d_i$.}
Agent $i$ does not directly choose her degree $d_i$ -- only the amount of time she spends inside her community, $t_i$. And $t_i$ influences the probability of realising each possible value of $d_i$, so we can write:
\begin{align}
    v_i^*(t_i) = \sum_{d=0}^D v_i^*(d_i) p(d| t_i).
\end{align}
By assumption, $p(d| t_i)$ is continuously differentiable in $t_i$ for all $d$ and all $t_i$. Therefore we have
\begin{align}
    \frac{d v_i^*(t_i)}{d t_i} = \sum_{d=0}^D v_i^*(d_i) \frac{ d p(d| t_i)}{d t_{i}}.
\end{align}
And finally, because $v_i^*(d_i)$ does not depend on $d_i$ when $n > n_0$ (i.e. $v_i^*(d_i) = \mathcal{V}_{i}$ for all $d_i$, we have
\begin{align}
    \frac{d v_i^*(t_i)}{d t_i} =0 \quad \text{for} \quad n > n_0
\end{align}
This follows from the fact that $\sum_{d=0}^D \frac{ d p(d| t_i)}{d t_{i}} = 0$ for all $d$ and all $t_i$ regardless of $n$ (as probabilities must sum to 1). For completeness, note that the choice of $t_i^{out}$ does not affect the network, and so trivially does not affect the second stage in any way.

\paragraph{Step 2: characterising first stage behaviour.} Write $u_i = I(t_i, \pi_i t_i^{out}) - c(t_i + t_i^{out}) + v_i(t_i)$. From above, we know that when $n> n_0$, $\frac{d v_i^*(t_i)}{d t_i} =0$. Therefore for $n > n_0$, choosing $t_i, t_i^{out}$ to maximise utility is the same as choosing $t_i, t_i^{out}$ to maximise $I(t_i, \pi_i t_i^{out}) - c(t_i + t_i^{out})$. By definition, the solution to this is $(\hat{t}_i, \hat{t}^{out}_i)$. Note that $(t_i^*, t_i^{out *}) = (\hat{t}_i, \hat{t}^{out}_i)$ is unique because $I(t_i, \pi_i t_i^{out})$ is strictly concave in both its arguments and $c(t_i + t_i^{out})$ is strictly convex. Up to this point, this is all subject to the assumption that \Cref{ass:degree_dist_convergence} holds.

\paragraph{Step 3: existence.} When \Cref{ass:degree_dist_convergence} holds, the resulting first stage choices (found in Step 2) satisfy \Cref{ass:degree_dist_convergence}. So an equilibrium exists. To see this, notice that first stage preferences only depend on an agent's group (high-skilled or low skilled). So $\hat{t}_i = \hat{t}_j$ and $\hat{t}^{out}_i = t_j^{out}$ whenever $i,j$ are in the same group. And a constant fraction $f$ are high-skilled (no matter how large $n$ becomes). So when $(t_i^*, t_i^{out *}) = (\hat{t}_i, \hat{t}^{out}_i)$, first stage choices follow a distribution with two point masses. As this remains the same for all $n > n_0$, we must have $\Lambda^{(n)}(t) \overset{d}{\to} \Lambda(t)$ as $n \to \infty$. (in fact we have something stronger: $\Lambda^{(n)}(t) = \Lambda(t)$ for all $n > n_0$). By \Cref{lem:technical_1}, this satisfies \Cref{ass:degree_dist_convergence}. \hfill \qed

\subsection*{A Lemma on group behaviour and network connectivity}
Before proving \Cref{rem:comp_stat_pi}, I first show that if there are a finite number of groups, with each group making up a fixed fraction of the whole community, then the measure of network connectivity that matters for public goods contributions is strictly increasing in the amount of time any one groups spends inside the community. 

\begin{lem}\label{lem:threshold_t_hat}
   Suppose there are finitely many groups, $h \in \mathcal{H}$, each making up a fraction $f_h$ of all agents. And suppose that for all $h$ we have $t_i = t_h$ for all $i \in h$. Then $\lim_{n \to \infty} \frac{1}{n} \sum_i d_i (d_i - 2) = \sum_h f_h Z(t_h)$ for some strictly increasing function $Z(\cdot)$.
\end{lem}

\noindent \emph{Proof.}
First, we can express a sum over all agents as a sum over agents within each group, then summed over groups. Second, for each group $h$ we have $t_i = t_h$ for all $i \in h$. So the fraction of agents in group $h$ with degree $d_i = d$ converges (as $n \to \infty$) to the probability that a randomly chosen agent in group $h$ has degree $d_i = d$. This is by a Law of Large Numbers. Each group is weighted by its size (the fraction of all agents in that group). Finally, multiply out.
\begin{align}
    \lim_{n \to \infty} \frac{1}{n} \sum_{i=1}^n d_i (d_i - 2) 
    &= \lim_{n \to \infty} \frac{1}{n} \sum_{h \in \mathcal{H}} \sum_{i\in h}^n d_i (d_i - 2) \\
    &= \sum_{h \in \mathcal{H}} f_h \sum_{d=0}^{D} d (d - 2) p(d | t_h) \\
    &= \sum_{h \in \mathcal{H}} f_h \left[ \sum_{d=0}^{D} d^2 P(d_i = d | t_h) - 2 \sum_{d=0}^D d P(d_i = d | t_h) \right]
\end{align}
The term in square brackets is simply the second moment of the distribution $p(d | t)$ minus twice the first moment. We have assumed that this is an increasing function of $t_h$. \hfill \qed 

\subsection*{Proof of Remark 1}
First, notice that $\hat{t}_i$ is strictly decreasing in $\pi_i$.  Therefore an increase in $\pi_H$ decreases $\hat{t}_i$ for the group of high-skilled agents. That $\sum_d d(d-2) \lambda_d$ is decreasing in $\pi_H$ then follows from \Cref{lem:threshold_t_hat}. That $x_i^*$ is weakly decreasing in $\pi_H$ then follows from this plus \Cref{prop:second_stage}.

\subsection*{Proof of Proposition 3}
\emph{First stage payoffs.} First, for any choices $(t_i, t_i^{out})$, $I(t_i, \pi_i t_i^{out}) - c(t_i + t_i^{out})$ is strictly increasing in $\pi_H$ for all high-skilled agents (as $\pi_i = \pi_H$). Re-optimisation can only further increase payoffs. So it follows from \Cref{prop:first_stage} that $I(t_i^*, \pi_i t_i^{out *}) - c(t_i^* + t_i^{out*})$ is strictly increasing in $\pi_H$ for all high-skilled agents. Second, it is clear that $I(t_i^*, \pi_i t_i^{out *}) - c(t_i^* + t_i^{out*})$ is unaffected by changes in $\pi_H$ for all low-skilled agents.

\emph{Second stage payoffs.} Recall from \Cref{rem:comp_stat_pi} that $\sum_d d(d-2) \lambda_d$ is strictly decreasing in $\pi_H$. But it follows from \Cref{prop:second_stage} that $x_i^*$ for all $i$ (and hence second stage payoffs) do not change unless $\sum_d d(d-2) \lambda_d$ crosses zero.\footnote{The term `crosses zero' is used somewhat loosely here. Formally, I take it to mean that  $\sum_d d(d-2) \lambda_d$ switches from being strictly positive to being weakly negative (or vie versa).} 

\emph{Putting things together.} So if $\sum_d d(d-2) \lambda_d$ does not cross zero, then only first stage payoffs are changing in $\pi_H$. In this case, payoffs are strictly increasing in $\pi_H$ for high-skilled agents and are invariant in $\pi_H$ for low-skilled agents. And if an increase in $\pi_H$ does cause $\sum_d d(d-2) \lambda_d$ to cross zero from above,\footnote{As $\sum_d d(d-2) \lambda_d$ is strictly decreasing in $\pi_H$ an increase in $\pi_H$ can only cause it to switch from strictly positive to weakly negative -- never the other way around.} 
then second stage payoffs strictly decrease. 

Finally, as first stage payoffs for high-skilled agents are continuously differentiable in $\pi_H$, there must be a small enough increase in $\pi_H$ such that the decrease in second stage payoffs from $\sum_d d(d-2) \lambda_d$ crossing zero from above outweighs the increase in first stage payoffs for high-skilled agents. \hfill \qed

\subsection*{Proof of Remark 2} 
\emph{First stage payoffs.} Let $I^H = \arg\max I(t_i, \pi_H t_i^{out}) - c(t_i + t_i^{out})$, and $I^L = \arg\max I(t_i, \pi_L t_i^{out}) - c(t_i + t_i^{out})$ denote first stage payoffs for high-skilled and low-skilled agents, respectively.\footnote{Formally, these are equilibrium first stage payoffs for $n$ sufficiently large. See the proof to \Cref{prop:first_stage}.}
It is clear that $I^H > I^L$.

\emph{Second stage payoffs.} Average second stage payoffs are the same for both groups, and depend only on whether or not we have $\sum_d d(d-2) \lambda_2 > 0$ (this follows from \Cref{prop:second_stage}). Denote them $v$. Then we have $v = v_1$ when $\sum_d d(d-2) \lambda_2 > 0$, and $v = v_2$ otherwise, where:
\begin{align}
    v_1 &= B(F(R-\epsilon)) + \sum_C \mathbb{P}(C_i = C) \cdot (R-C) \cdot \mathbf{1}\{C<R\} \\
    v_2 &= B(F(0)) + \sum_C \mathbb{P}(C_i = C) \cdot -C \cdot \mathbf{1}\{C\leq 0\}
\end{align}
for $\epsilon>0$ sufficiently small.\footnote{We know from \Cref{prop:second_stage} that all agents with $C_i < R$ choose $x_i^* = 1$ (and everyone else chooses $x_i^*$) when $\sum_d d(d-2) \lambda_2 > 0$, and that all agents $C_i \leq 0$ choose $x_i^* = 1$ (and everyone else chooses $x_i^*$) when $\sum_d d(d-2) \lambda_2 \leq 0$. The first term is the benefits from public goods when people behave like this. The second term (with the summation) is the benefit that agents get from contributing to public goods and receiving any reward, weighted by the fraction of agents with each value of $C_i$.}
Then group-average payoffs are $u^H = I^H + v$ and $u^L = I^L + v$, for the high-skilled and low-skilled groups, respectively.
%
\emph{Gini Coefficient.} Recall that:
\begin{align}
    Gini = \frac{1}{2 \mu} \sum_g \sum_{g'} \phi (g) \phi(g') |u_g - u_{g'}|
\end{align}
where $\mu = \sum_g \phi(g) \cdot u_g$, and $\phi(g)$ is the fraction of agents who are in group $g$. We have two groups, $H$ and $L$, with $\phi(H) = f$ and $\phi(L) = 1-f$. Therefore, we have:
\begin{align}\label{eq:gini_result}
    Gini = \frac{1}{2 \mu} 2 f (1-f) (I^H - I^L).
\end{align}
where $\mu = f(I^H - I^L) + I^L + v$. 
We know that $I^H$ is strictly increasing $\pi_H$ and $I^L$ is invariant in $\pi_H$ (\Cref{prop:welfare}). So when $\pi_H$ does not affect public good provision, we have that between-group inequality is increasing in $\pi_H$. This is because 
\begin{align}
    \frac{d Gini}{d I^H} > 0
\end{align}
When $\pi_H$ \emph{does} affect public good provision, it must be that an increase in $\pi_H$ reduces public good provision (\Cref{rem:comp_stat_pi}) -- and so reduces $v$ from $v_1$ to $v_2$. This reduces $\mu$ without affecting any other part of \Cref{eq:gini_result}, which increases between-group inequality. \hfill \qed




\subsection*{Proof of Proposition 4}
First notice that conditional on $\sum_d d (d-2) \lambda_d$, the utility of any given low-skilled agent is invariant in the number of other low-skilled agents. Second, $\sum_d d (d-2) \lambda_d$ is strictly decreasing in $f$. This follows from \Cref{lem:threshold_t_hat} and the fact that $t_i^* > t_j^*$ whenever $i$ is high-skilled and $j$ is low-skilled. Finally, utility strictly falls when $\sum_d d (d-2) \lambda_d$ `crosses zero' from above. \hfill \qed



\cleardoublepage
\begin{center}
\section*{Online Appendix}    
\end{center}
\newpage
\section{The role of a technical assumption}\label{OA:tech_ass}
My model imposes a technical assumption that $\mathbb{E}[d^2 | t_i] - 2 \mathbb{E}[d | t_i]$ is strictly increasing in $t_i$. What this does is ensure that the precise measure of network connectivity that matters in my setting (which is $\sum_{d} d (d - 2) \lambda_d$) is increasing in $t_i$ (at least if a positive fraction of agents increase their $t_i$). So I am \emph{assuming} that when people spend more time inside their community, the social network in said community becomes `better connected' according to the metric that matters.

Suppose I were to relax this assumption, and instead only require that more time spent inside the community leads to more links in the first order stochastic dominance sense (i.e. that for all $t'_i > t_i$ we have $\sum_{d=0}^{D'} P(d_i = d | t'_i) \geq \sum_{d=0}^{D'} P(d_i = d | t_i)$ and for all $D' \leq D$, with at least one strict inequality). Then network connectivity could \emph{fall} when people spend more time inside the community. From a technical standpoint, this is due to the impact agents with exactly one friend have in random networks. Recall that the each agent's degree, $d_i$, is determined before the network is realised. So increasing the degree of some agents is not quite the same as adding links to an already realised network. 

To see how this works, consider a community where some small fraction of agents have degree 4, but most have degree $0$. All degree-4 agents must then be connected to other degree-4 agents.\footnote{In effect we have a network where everyone has 4 friends, and separately some redundant agents who play no role. When thinking about a network, it is possible to put agents with degree zero aside and consider them separately. Note that I have ignored the technical assumption that $\lambda_1 > 0$ here. This is for convenience of exposition only.} The probability that all degree-4 agents are part of a single giant component is one in the limit $n \to \infty$. This is because they are forming links at random---so the chance there is group of agents where nobody has any links outside of that group is vanishingly small. 

\begin{figure}[h!]
    \centering
    \begin{subfigure}{0.49\textwidth}
        \centering
        \includegraphics[width=\linewidth]{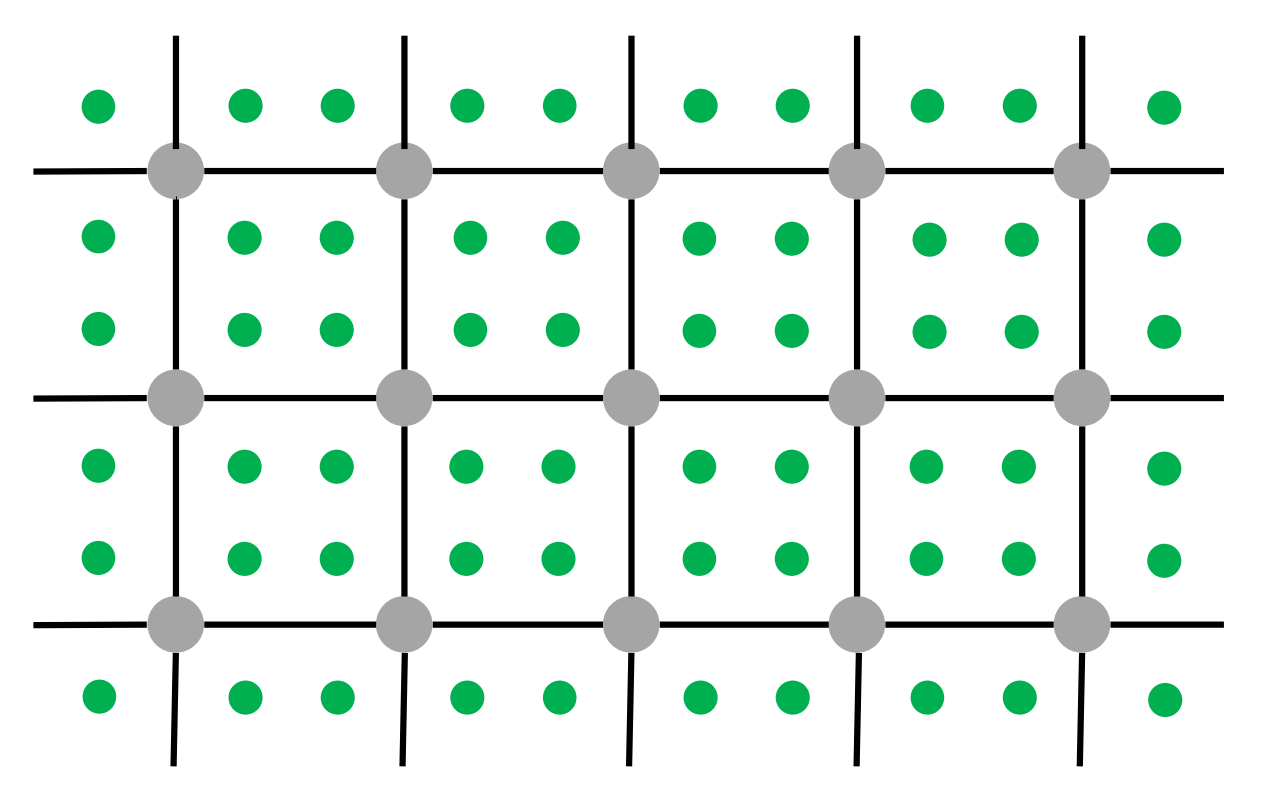}
        \caption{}
        \label{fig:tech_ass_A}
    \end{subfigure}
    \hfill
    \begin{subfigure}{0.49\textwidth}
        \centering
        \includegraphics[width=\linewidth]{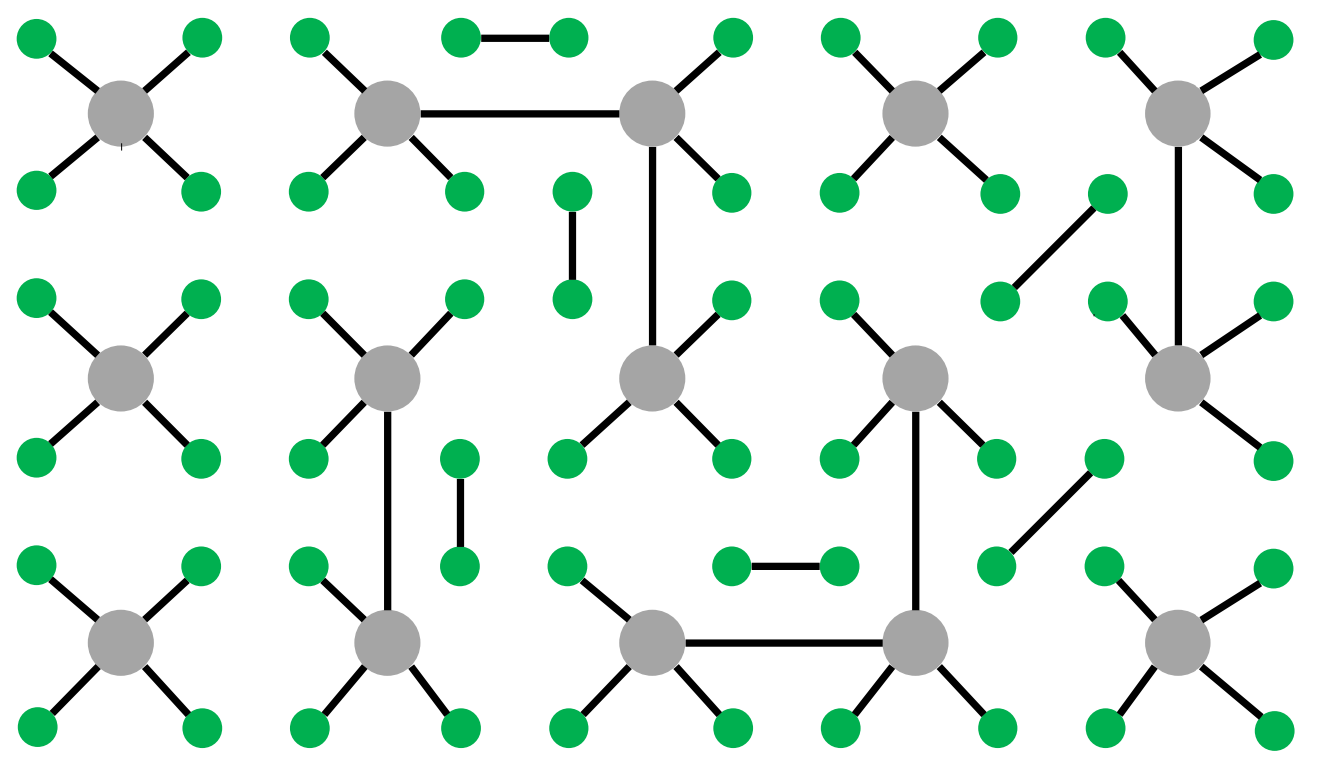}
        \caption{}
        \label{fig:tech_ass_B}
    \end{subfigure}

    \caption{A stylised illustration of how increasing degrees from 0 to 1 can destroy a giant component. Panel (A): gray nodes have degree 4, green nodes have degree 0. One large component containing all degree-4 nodes. Panel (B): gray nodes have degree 4, green nodes have degree 1. Many small components.}
    \label{fig:tech_ass}
\end{figure}

This changes if the degree-0 agents all make a friend: increasing their degree to 1 (for example, by spending more time inside their community). The now-degree-1 agents make up the majority of the community, so degree-4 agents are likely to form links with degree-1 agents. The degree-4 agents are then likely to end up at the centre of a star, connected only to degree-1 agents.\footnote{Even when degree-4 agents do form links with each other, (indirectly) connected groups of degree-4 agents will eventually have no links outside of the group, because of their links to degree-1 agents (who by definition will never have links outside of the group).}
This special role of agents with degree 1 -- acting as a `dead end’ in the flow of information -- is what allows increases in the degree (specifically, from 0 to 1) of some agents to stop a giant component from forming. However, this is driven by the random link formation. In a world with homophily on degree -- where agents with lots of friends are more likely to connect with others who also have lots of friends -- this feature would not be present. High degree agents would still be likely to connect with one another, even in a community with many degree-1 agents.

This is why I impose this technical assumption: I want to rule out an effect driven by the absence of homophily -- because homophily is a ubiquitous feature of real networks \cite{jackson2008, jackson2023dynamics}. The technical assumption requires that when an agent spends more time inside her community, the probability that this leaves her with a degree of exactly 1 does not rise by too much relative to the increase in the probability of being left with higher degrees.


\newpage
\section{Multiple Communities}\label{OA:multiple_comms}
Suppose that the model is exactly as in \Cref{sec:model} with the following adjustments:

\begin{itemize}
    \item[(1)] there are now $M$ different communities, $m \in \{1,..,M\}$, each with $n$ agents. In an abuse of notation we let $m$ denote both the community's index and the set of agents it contains.
    
    A community is defined as before: for each $m$ separately, the identities of all $i \in m$ are common knowledge amongst $i \in m$. Importantly, any pair of agents $i,j$ with $i \in m$ and $j \notin m$ do not know each other's identity.

    \item[(2)] All parameters and functions that were common to all agents in \Cref{sec:model} are indexed $m$ -- allowing them to vary across communities (while remaining common to all agents within a community $m$)

    \item[(3)] The observation, reward and social networks all form separately for each community $m$. Formally, the random network processes become $\mathcal{G}_m^o((d_i^o)_{i \in m})$.
\end{itemize}

Given this, it is clear that each community is playing the game separately. Decisions in community $m$ can never affect the incentives or payoffs of agents in any community $m' \neq m$. But as we allow the parameters (notably $\pi_H$ and $p(d|t)$) to differ across communities, in equilibrium we can have some communities with high public good provision and some with low provision. This is simply because with different parameter values comes different behaviour. 

An immediate implication of this is that a common shock to all communities that makes it more attractive to interact outside of one's own community (i.e. a shift from $\pi_{Hm}$ to $\pi'_{Hm} = \pi_{Hm} + k$ for some $k>0$ and for all $m$) can have different effects on different communities. Some may experience a large fall in public good provision, while others see no change.
This independent behaviour by different communities is driven by the fact that information about an agent $i$'s behaviour can only spread between others who both know $i$'s identity, and we have defined a community as the group of people who all know $i$'s identity (see \Cref{sec:interpretation} for discussion).

Here, we have shown that the model extends easily to a setting with many communities, when agents are only a member of one community. Having agents be members of more than one community at the same time complicates matters significantly. We only provide some very limited results in this direction, and discuss some of the challenges it creates.

\subsection{Multiple membership}
When an agent $i$ is a member of multiple different communities, all agents in all of her communities know her identity. This means that information about $i$'s behaviour can flow both within each of her communities and \emph{between} the different communities she is part of.\footnote{As before, information about $i$'s behaviour cannot flow through (or to) communities that she is not part of. It remains the case that only links between agents who both know $i$'s identity matter.} 
This means that the interactions between these communities now matter. To give a flavour of what can happen, we consider the case where an agent $i$ is a member of two different communities, $A$ and $B$.

Additionally, we suppose that the public goods are specific to the community. Specifically, $i$ chooses whether or not to provide some public good(s) in each of her communities separately: $x_{iA} \in [0,1]$ and $x_{iB} \in [0,1]$. And all of the observers and rewarders for action $x_{iA}$ are in community $A$ (and similarly for action $x_{iB}$). This is motivated by the earlier assumption that only people within a given community have a shared interest in a particular community-level public good. To focus on the interesting case, also assume $C_{iA} < R$ and $C_{iB} < R$.
In this setting, $i$'s behaviour will depend on both the within-community networks, $G_A(i)$ and $G_B(i)$, and the between-community network, $G_{AB}(i)$. The between-community network, $G_{AB}(i)$, is a realisation of a \emph{bipartite} configuration model. This selects a unweighted and undirected bipartite network uniformly at random from those with the degree sequences $\mathbf{d_{AB}}^{(n)}$ and $\mathbf{d_{BA}}^{(n)}$. 
The number of links, $d_{AB, j}$, an agent $j$ in community $A$ forms with agents in community $B$ is a draw from a Probability Mass Function $p_{out}(d^{out} | t^{out})$. And similarly for agents in community $B$. This creates one degree sequence for each community, denoted $\mathbf{d_{AB}}^{(n)}$ and $\mathbf{d_{BA}}^{(n)}$, respectively.

Suppose that $G_A(i)$ does not have a giant component, but $G_B(i)$ does. Then by \Cref{prop:second_stage}, we have $x_{iB}^* = 1$ for large enough $n$. But the equilibrium choice of $x_{iA}$ depends on the between-community network $G_{AB}(i)$. In the special case when $G_{AB}(i)$ is an empty network, then clearly $x_{iA}^* = 0$ for large enough $n$. But even a modest degree of connectedness between the two communities can change $i$'s behaviour. All agents in each community having at least one between-community link is sufficient to make agent $i$ provide public goods in community $A$. This does not depend on how sparse $G_A$ is -- it can even be an empty network.

This is because the probability that at least one observer from community $A$ and at least one rewarder from community $A$ both have a link to someone in the giant component of community $B$ tends to 1 as $n \to \infty$. So the probability that $i$ gets a reward for choosing $x_{iA} = 1$ tends to 1 as $n \to \infty$. Hence $x_{iA}^* = 1$ for large enough $n$ (because $i$ chooses $x_{iA}^* = 1$ whenever $p_i(\text{reward}_A) \geq C_{iA} / R$).

\newpage
\section{Generalised Information Spreading}\label{OA:gen_info}
This section formalises the discussion in \Cref{subsec:gen_info}. First, it considers the case where some fraction of agents never pass on any information, and those who do only pass information with some exogenous probability. I show that the qualitative result does not change, but we can no longer give an explicit function for the threshold on network connectivity. Second, it considers a setting where passing information is a strategic decision (as is providing rewards). This takes an alternative view to the main model -- which views gossip and rewards as merely incidental, where agents do not go out of their way to do them.

\subsection{Non-strategic frictions}
\paragraph{Adjusting the model.} The model is as in \Cref{sec:model} with the following changes:
(1) with i.i.d probability $Q \in [0,1)$, each agent is `silent'. When `silent', an agent never passes information to her friends. 
(2) remaining agents are willing to pass information to each of their friends in the social network with i.i.d. probability $q > 0$.\footnote{Everything else -- agents, actions, timing, network formation, rewards, preferences, information and the solution concept -- are unchanged. I do not repeat these parts of the model here} 

Passing information requires mutual consent: we say a link is `active' if and only if agents at both ends of a link are willing to pass information. So agents who are path connected (in the social network) by active links to some $j \in \mathcal{N}_i^o$, all learn $i$'s action.\footnote{Two agents $i,i'$ are path connected by active links if there exists a sequence of links $G_{i,1},...,G_{k,i'}$ such that every link in the sequence is active.}

\paragraph{How this generalises the main model.}
The second direction is to relax the stark assumption that information flows frictionlessly. We do this in two related ways. \textbf{(a)} Information about $i$'s action is only communicated across a link with some exogenous probability. This is a reduced-form way of capturing frictions in information transmission, and might reflect the fact that this information is incidental gossip, not the motivating reason for establishing links. \textbf{(b)} Some agents may be `silent' (for some exogenous reason), and \emph{never} pass on information about $i$'s action to their friends. In the context of using information to hold others in a community accountable, this silence could be viewed as a form of anti-social behaviour. One restriction I do impose on these frictions is that they are not connected to the number of friends an agent has. From a technical standpoint, adding these frictions amounts to randomly deleting links and nodes respectively from the main network. And therefore the assumption that information flows non-strategically remains important for tractability.

\paragraph{The generalised result.}
Before stating the generalised version of \Cref{prop:second_stage}, it is useful to define a term that will act as the analogy to $\sum_d d(d-2) \lambda_d$, but is a little more cumbersome. Let

\begin{align}
    \chi := (1-Q) \sum_{d=0}^{\infty} d (d - 1) \tilde{\lambda}_d - \sum_{d=0}^{\infty} d \tilde{\lambda}_d,
\end{align}
where 
$\tilde{\lambda}_{d'} = \sum_{d \geq d'} \mathbb{P}(\text{Bi}(d, \psi) = d') = \binom{d}{d'} \psi^{d'} (1 - \psi)^{d - d'} \lambda_{d}$, and $\psi := (1-q)^2$. With this in place, we can now state the result.

\begin{prop}\label{prop:generalised_result}
Fix some $q, Q$, and suppose $\{ \mathbf{d}^{(n)} \}_{n \in \mathbb{N}}$, is such that $\lim_{n \to \infty} \frac{1}{n} |\{i : d_i = d\}| = \lambda_d \geq 0$ for all $d$. Then there exists an $n_0$ such that for all $n > n_0$,

(i) \ if $\chi > 0$, then $x_i^* = 1$ if $C_i \leq 0$ and $x_i^* = 0$ otherwise,
    
(ii) if $\chi \leq 0$, then $x_i^* = 1$ if $C_i < R$ and $x_i^* = 0$ otherwise.  
\end{prop}

The step function remains. There is a threshold in network connectivity above which agents behave `well' (or at least as well as dense networks could every get them to behave). And below which agents behave `badly'. The intuition for this result is unchanged compared to \Cref{prop:second_stage}. The driving force is still that the network spreads the information needed to hold agents accountable for their behaviour.

\paragraph{How the proof extends.} The proof strategy is also little changed, although it requires some extra steps. What matters for hold agents to account is not links \emph{per se}, but `active' links -- links across which information flows. So we are interested in the component structure of the network of active links, rather than of all links. Trivially, some probability that information does not flow along individual links (i.e. are `inactive') will raise the connectivity of the initial network that is required to get a giant component in the network of active links. But it does not alter the result that the giant component appears at \emph{some} threshold of connectivity. 

From a technical standpoint, stochastic information flow amounts to random deletion of nodes (agents) and links from the social network to construct the `active' social network. It turns out that random deletion of links after the formation of a random network is equivalent to making an adjustment to the degree distribution \emph{before} the network forms. 
Random deletion of nodes cannot be dealt with in exactly the same way. For this, we rely on existing results from graph theory showing that for a network with a giant component, there is a critical probability of node deletion probability at which the giant component disappears.

\paragraph{Proof.}
\emph{Step 1: Active Network.} For the purposes of learning about $i$'s action, it is clear that only `active' links in the gossip network $G$ matter, where a link $G_{j \ell}$ is `active' if and only if information passes along it (i.e. if $\ell$ learns about $i$'s action, conditional on $j$ having learned about it). Define $G^{act}$ to be the sub-network of $G$ consisting of only active links. $G^{act}$ can clearly be constructed in three steps: (1) take a realisation of $\mathcal{G}(\lambda_0, ... , \lambda_D)$, (2) keep each link with probability $(1 - q)^2$,\footnote{If each agent deletes each of her links with probability $q$, then the probability it is kept is equal to the probability that \emph{both} of the two agents who are party to the link keep it. Each keep independently with probability $(1-q)$.} 
then (3) remove all of a node's links with probability $Q$. \\

\noindent \emph{Step 2: Giant component of $G^{act}$ is necessary and sufficient.} This is by the same logic as the proof to \Cref{prop:second_stage} (see especially, step 1 and step 2). \\

\noindent \emph{Step 3: Giant component of $G^{act}$.} A configuration model, $\mathcal{G}((\lambda_d)_{d=0}^{\infty})$ followed by deleting links uniformly at random with probability $(1-q)^2$, is itself a configuration model (with different parameters), $\mathcal{G}((\Tilde{\lambda}_d)_{d=0}^{\infty})$ (see for example, \citet[esp. S.4 and S.6]{bollobas2015old} \citet[S.1.1]{fountoulakis2022percolation}). We can calculate the $\Tilde{\lambda}_d$'s straightforwardly. 

As $n$ becomes large, a fraction tending to $\lambda_d$ will initially have degree $d$. The probability that each agent with degree $d$ ends up with degree $d'$ after the random deletions is $b_{d, d'}(\psi)$, where
\begin{align*}
    b_{d, d'}(\psi) = \mathbb{P}(\text{Bi}(d, \psi) = d') = \binom{d}{d'} \psi^{d'} (1 - \psi)^{d - d'},
\end{align*}
and we are using $\psi := (1-q)^2$. By a standard Law of Large Numbers, the fraction of agents `transitioning' from degree $d$ to $d'$ due to the random deletions tends to $b_{d, d'}(\psi)$ as $n$ becomes large. With a fraction $\lambda_d$ agents starting with degree $d$ (before deletions), the fraction who moved from degree $d$ to $d'$ is simply $\lambda_d \cdot b_{d, d'}(\psi)$. So these random link deletions induce a new degree distribution $(\tilde{\lambda}_0,...,\tilde{\lambda}_D)$, with
\begin{align*}
    \tilde{\lambda}_{d'} = \sum_{d \geq d'} b_{d , d'}(\psi) \lambda_{d}.
\end{align*}

So we can now consider whether $\mathcal{G}((\tilde{\lambda}_0, ... , \tilde{\lambda}_D))$ has a giant component after randomly selecting each agent with probability $Q$ and deleting all of their links.
\cite[Theorem 3.5]{janson2009percolation} considers exactly this (often called  `site percolation' in the graph theory literature).\footnote{Formally, he considers the random deletion of agents and all of their links. But once all of an agent's links are removed, they can never learn about $i$'s action or tell anyone else. So it does not matter for the analysis whether the agent is removed from the network or left as a node with degree zero.}  
We will restate a simplified version of this result (the original result allows the probability a node is deleted to depend on its initial degree, but we do not allow that), and only restate the necessary parts. 

\begin{thm*}[Janson (2009), Theorem 3.5]
Consider the site percolation model $G^*(n,d)_{(1-Q),v}$ and suppose that Condition 2.1 holds and that $0 \leq (1-Q) \leq 1$; suppose further that there exists $d \geq 1$ such that $\lambda_d > 0$. Then there is w.h.p. a giant component if and only if
\begin{align}\label{eq:janson_GC_condition}
    (1-Q) \sum_{d=0}^{\infty} d (d - 1) \tilde{\lambda}_d >  \sum_{d=0}^{\infty} d \tilde{\lambda}_d
\end{align}
\end{thm*}

This result tells us exactly when $G^{act}$ has a giant component with high probability, and when it does not have a giant component with high probability. \hfill \qed

\subsection{Strategic information spreading and reward provision}

\paragraph{Adjusting the model.} 
In the interest of simplicity, $i$ simply chooses (1) whether or not to gossip to her friend $j$ about anyone's action, and (2) whether or not to provide a reward when called on to do so (conditional on learning of public good provision by the recipient of the reward). I assume that she cannot condition her decision on the identity of the agent whose action she knows about. 

Formally, each agent $i$ chooses $y_i \in \{0,1\}$ (whether or not to reward) and $g_{ij} \in \{0,1\}$ for each $j$ such that $G_{ij} = 1$ (whether or not to pass on information). Both providing a reward and passing on information are themselves seen as public goods. So all $j \in \mathcal{N}_i^o$ observe the decision, and $i$ will receive a reward $\Tilde{R} > 0$ if at least one of her rewarders, $j \in \mathcal{N}_i^r$, learns of her public good provision \emph{and} that rewarder has chosen to provide rewards when called upon. 

Each agent also has a cost $\Tilde{C}_{ij} \in \mathbb{R}^+$ of gossiping to $j$ and a cost $\Tilde{C}_i \in \mathbb{R}^+$ of providing a reward, which are draws from some Probability Mass Functions that do not depend on $n$, $\Tilde{F}(\cdot)$ and $\Tilde{\Tilde{F}}(\cdot)$, respectively. So $i$ receives a payoff of:
\begin{align*}
    \Tilde{v}_i = p_i(\text{reward}) \Tilde{R} y_i - \Tilde{C}_i y_i + \left( \sum_{j \in \mathcal{N}_i(G)} p_i(\text{reward}) \Tilde{R} g_{ij} - \Tilde{C}_{ij} g_{ij} \right)
\end{align*}

Assume that agent $i$ takes an action (whether that is transmitting information or providing a reward) when indifferent. Let $\Tilde{F}(\cdot)$ be the Cumulative Distribution Function of $\Tilde{C}_i$. For simplicity, also assume that $\tilde{F}(0)>0$.

\emph{Some notation.} Recall that the realised gossip network is denote $G$. In a small abuse of notation we let $G[0]$ denote the subset of the realised gossip network, containing only the links where $\tilde{C}_{ij} = 0$. These are special because agents will transmit information along these links regardless of what else happens. Also, we will talk about a network `having a Giant Component' (or not having one). We have already discussed in the main text when one exists. Note that realising the costs $\Tilde{C}_{ij} \in \mathbb{R}^+$ before the network $G$ is important here. It means that if $\Tilde{\Tilde{F}}(\cdot)$ has an atom at zero, then...

\paragraph{The result.}

\begin{prop}
Define two patterns of behaviour: \\
(a) $y_i^* = 1$ if and only if $\Tilde{C}_i \leq \Tilde{R}$, and $g_{ij} = 1$ if and only if $\Tilde{C}_{ij} \leq \Tilde{R}$, \\
(b) $y_i^* = 1$ if and only if $\Tilde{C}_i = 0$, and $g_{ij} = 1$ if and only if $\Tilde{C}_{ij} = 0$. \\

Suppose $\{ \mathbf{d}^{(n)} \}_{n \in \mathbb{N}}$, is such that $\lim_{n \to \infty} \frac{1}{n} |\{i : d_i = d\}| = \lambda_d \geq 0$ for all $d$. Then there exists an $n_0$ such that for all $n > n_0$,

\noindent (i) \ \ If $G$ has a giant component, but $G[0]$ does not: there are two Nash equilibria, (a) and (b), \\
(ii) \ If $G[0]$ has a giant component: (a) is the unique Nash equilibrium, \\
(iii) If that $G$ does not have a giant component: (b) is the unique Nash equilibrium.
\end{prop}

\paragraph{Proof}
Call a link $G_{jk}$ \emph{active} if $g_{jk} = 1$ (i.e. $j$ chooses to tell $k$ about $i$'s behaviour) and \emph{inactive} otherwise. So let $G(\text{active})$ be the network of active links. Formally, $G(\text{active})$ is such that $G(\text{active})_{jk} = 1$ if $G_{jk} = 1$ AND $g_{jk} = 1$, and $G(\text{active})_{jk} = 0$ otherwise. 

Information flows only along links in $G(\text{active})$. So this is the relevant network for information transmission. And it follows from the argument in \Cref{prop:main_result} that if $G(\text{active})$ has a Giant Component, then some positive fraction of all of $i$'s `rewarders' will learn of her public good provision.

It is clear that agents $j$ for whom $\Tilde{C}_j = 0$ will always choose $y_j = 1$ (as we assumed they provide when indifferent). This means if they learn of public good provision by $i$, and they are in a position to reward $i$ (that is, $j \in \mathcal{N}_i^r$) they will provide the reward when called upon. And recall that we assumed $\tilde{F}(0)>0$, so a positive fraction of all agents provide rewards in this way.

Putting these two pieces together, if $G(\text{active})$ has a Giant Component, then at least one of $i$'s `rewarders' will be willing to provide the reward---so $i$ will be rewarded for public good provision. And if $G(\text{active})$ does not have a Giant Component, $i$ will never get a reward---because no `rewarder' will learn of her action (exactly as in \Cref{prop:second_stage}).

Finally, recall that for agents with a cost parameter that is positive but less than $\Tilde{R}$, they prefer to take the action if they will get a reward. And they prefer not to take the action if they won't. For agents with costs of zero, or above $\Tilde{R}$, their best responses do not depend on the network at all.

With this, we then examine Best Response functions.
\begin{align*}
    BR_{j}(G(\text{active}), y_{-j}) = 
    \begin{cases}
        y_{j} = 1 &\text{ if } \Tilde{C}_{j} = 0, \\
        y_{j} = 1 &\text{ if } \Tilde{C}_{j} \in (0,\Tilde{R}] \text{ and } G(\text{active}) \text{ has a giant component, }\\
        y_{j} = 0 &\text{ if } \Tilde{C}_{j} \in (0,\Tilde{R}] \text{ and } G(\text{active}) \text{ does not have a giant component, }\\
        y_{j} = 0 &\text{ if } \Tilde{C}_{j} > \Tilde{R}. \\
        g_{jk} = 1 &\text{ if } \Tilde{C}_{jk} = 0, \\
        g_{jk} = 1 &\text{ if } \Tilde{C}_{jk} \in (0,\Tilde{R}] \text{ and } G(\text{active}) \text{ has a giant component, }\\
        g_{jk} = 0 &\text{ if } \Tilde{C}_{jk} \in (0,\Tilde{R}] \text{ and } G(\text{active}) \text{ does not have a giant component, }\\
        g_{jk} = 0 &\text{ if } \Tilde{C}_{jk} > \Tilde{R}. \\
    \end{cases}
\end{align*}

We know from earlier that the existence of a giant component is binary in a configuration model---it either exists or it doesn't. And we can see above that it is the only feature of the network that matters for best responses. 

Further, there are only two different best response action profiles: $y_{j}^* = 1 \iff : \Tilde{C}_{j} \leq \Tilde{R}, g_{jk}^* = 1 \iff : \Tilde{C}_{jk} \leq \Tilde{R}$ (for when there is a giant component) and $y_{j}^* = 1 \iff : \Tilde{C}_{j} = 0, g_{jk}^* = 1 \iff : \Tilde{C}_{jk} = 0$ (for when there is not). So there are the only candidates for equilibria. We now examine when there are or are not equilibrium action profiles. In an abuse of notation, we will let $G(\Tilde{C}_{jk} \leq X)$ denote the network where a link $G_{jk}$ is active if and only if $\Tilde{C}_{jk} \leq X$.

First, suppose that $G(\Tilde{C}_{jk} \leq 0) = G[0]$ has a giant component. Then \emph{(i)} $G(\Tilde{C}_{jk} \leq 0)$ is not a best response to itself---as agents with $\Tilde{C}_{jk} \in (0,\Tilde{R}]$ want to deviate---and so it not an equilibrium. And \emph{(ii)} $G(\Tilde{C}_{jk} \leq \Tilde{R})$ must have a giant component---as adding more links to a network can never remove a giant component. So then $G(\Tilde{C}_{jk} \leq \Tilde{C})$ \emph{is} a best response to itself, and so is the unique equilibrium.

Second, suppose that $G(\Tilde{C}_{jk} \leq \Tilde{R})$ does not have a giant component. Then \emph{(i)} $G(\Tilde{C}_{jk} \leq \Tilde{R})$ is not a best response to itself---as agents with $\Tilde{C}_{jk} \in (0,\Tilde{R}]$ want to deviate---and so it not an equilibrium. And \emph{(ii)} $G(\Tilde{C}_{jk} = 0)$ cannot have a giant component either---as removing more links from a network can never create a giant component. So then $G(\Tilde{C}_{jk} = 0)$ \emph{is} a best response to itself, and so is the unique equilibrium.

Finally, suppose that $G(\Tilde{C}_{jk} \leq \Tilde{R})$ has a giant component, but $G(\Tilde{C}_{jk} = 0)$ does not.\footnote{The final combination, where $G(\Tilde{C}_{jk} \leq \Tilde{R})$ does not have a giant component but $G(\Tilde{C}_{jk} = 0)$ does cannot exist. We have shown above that the existence of a giant component in $G(\Tilde{C}_{jk} = 0)$ implies the existence of one in $G(\Tilde{C}_{jk} \leq \Tilde{R})$.} 
Here, both are best responses to themselves, and so both are equilibria.

The equilibrium choices of $y_j$ then follow straightforwardly. The fact that $\Tilde{F}(0) > 0$ by assumption means that existence of the giant component is both necessary and sufficient for agents to get their reward for public good provision. Choices of $y_j$ for higher cost agents (those with $\Tilde{C}_j > 0$) do not end up mattering for others' incentives. \hfill $\square$

\newpage
\section{When agents have a material impact on network connectivity}\label{OA:coord_extension}
In this extension, I now have equal-sized groups of agents -- called \emph{households} -- coordinate their choices of $t_i$. This means that, unlike in the main model, agents’ first stage choices continue to have a material impact
on the network structure as $n$ becomes large.

\subsection{Adjusting the model}\label{sec:model_extended}
The model is as in \Cref{sec:model}, with the following adjustments:
\begin{itemize}
    \item[(1)] There are $H$ households $h \in \{1, ... , H\}$. Each agent $i$ is a member of exactly one household. All households are the same size, and contain only agents of one type (either high-skilled or low-skilled). In an abuse of notation, let $h$ denote both a household's index and the set of agents it contains.

    \item[(2)] One agent $i$ in each household $h$ is selected as a `head', and chooses $t_h$. All agents $i \in h$ must then adopt $t_i = t_h$. Also assume that the `head' of all households has $C_i > R$. 

    \item[(3)] After choosing $t_i = t_h$, and realising their degrees, $d_i$, a (small) fraction $\xi$ of agents who have at least two friends (i.e. $d_i \geq 2$) are selected at random and gain one extra friend. $\xi$ is a random draw from some continuous and atomless CDF, with support $[-\bar{\xi}, \bar{\xi}]$. When $\xi < 0$, this means that a fraction $-\xi$ \emph{lose} one friend. This induces a new degree sequence $(\tilde{d}_i)_{i=1}^n$, and a new fraction of agents with degree $d$, $\tilde{\lambda}_d$. For all $\bar{\xi}>0$, let the CDF be $Pr(\xi \leq z) = \widehat{F}(z / \bar{\xi})$, with $\widehat{F}(z / \bar{\xi}) = 0$ for all $z \leq -1$, $\widehat{F}(z / \bar{\xi}) = 1$ for all $z \geq 1$. This demands that the CDF `looks the same' as $\bar{\xi}$ changes. 

    This addition to the model is for technical reasons, and helps avoid non-existence of best responses in the first stage of the game due to an open interval problem. I consider what happens as the shock becomes small: i.e. $\lim \bar{\xi} \to 0$.
\end{itemize}

The decisions, preferences, network formation, information flow, information and timing are otherwise identical to \Cref{sec:model}. It is clear that conditional on first stage behaviour, this adjustment has no impact on the second stage. \Cref{prop:second_stage} holds unchanged. So I focus on the first stage, beginning with best responses.

\subsection{Best Responses}
Because the number of households is fixed, each household has a non-vanishing impact on overall network connectivity, even as $n$ grows large. So overall network connectivity is strictly increasing in $t_h$. In turn, this means that, for a given set of choices by other households, there is a unique value of $t_h$ such that the critical threshold in network connectivity is reached. Denote it $t_h^{crit}(t_{-h})$. Given the result from \Cref{prop:second_stage}, second stage payoffs are `high' (and the social fabric of the community is intact) when $t_h \geq t_h^{crit}(t_{-h})$, and is `low' otherwise. Let $\Delta v>0$ denote the utility loss of moving from `low' to `high' second stage payoffs.

First stage payoffs are as before. For any given $t_h$, there is a unique $t_h^{out}$ that maximises first stage payoffs. So it is convenient to write first stage payoffs as $w_h(t_h) = \max_{t_h^{out}} I(t_h, \pi_h t_h^{out}) - c(t_h + t_h^{out})$. As before, this is maximised at $t_h = \hat{t}_h$ (the same value as in \Cref{prop:first_stage}). Choosing $t_h > \hat{t}_h$ reduces first stage payoffs. But it may lead to `high' second stage payoffs. There will be some maximum value of $t_h$ that the household will be willing to choose in order to obtain the `high' second stage payoffs. Denote this $t^{max}_h$, where $w_h(t^{max}_h) = \Delta v + w_h(\hat{t}_h)$, and $t^{max} \geq \hat{t}$.

Having set out this extra notation, best responses are now straightforward.

\begin{rem}[Best Responses]\label{rem:best_response}
\begin{align}\label{eq:BR_HH}
    \lim_{n \to \infty} BR_h(t_{-h}) = 
    \begin{cases}
        t^{crit}_h  & \text{ if } t^{crit}_h \in [\hat{t}_h , t^{max}_h] \\
        \hat{t}_h & \text{ otherwise. }
    \end{cases}
\end{align}
\end{rem}

If $\hat{t}_h$ will lead to `high' second stage payoffs, then there is no reason to deviate from there. If choosing some $t_h > \hat{t}_h$ is needed to get `high' second stage payoffs (i.e. $t^{crit}_h > \hat{t}_h$), then the household will only do so if the gain in the second stage more than offsets the loss in the first stage (i.e. $t^{crit}_h \leq t^{max}_h$). And in that case they would only ever choose $t_h = t^{crit}_h$ -- because deviating further from $\hat{t}$ would further reduce first stage payoffs.

\paragraph{Proving \Cref{rem:best_response}.} The proof to \Cref{rem:best_response} involves a pair of technical steps. First, I show that the degree sequence after the shock $\xi$ is equal to the sum over all households of some increasing function of $t_h$, plus some term proportional to $\xi$. This means that whether or not network connectivity is above the critical threshold could, in principle, depend on the realisation of $\xi$. Second, I show that when $\xi$ is drawn from a sufficiently narrow distribution, this never happens. More precisely, households will never want to make choices that leaves network connectivity in a position where the realisation of $\xi$ can affect whether or not network connectivity is above the critical threshold. 

\begin{figure}[ht!]
    \centering
    \begin{subfigure}[b]{0.475\textwidth}
        \centering
        \includegraphics[width=\textwidth]{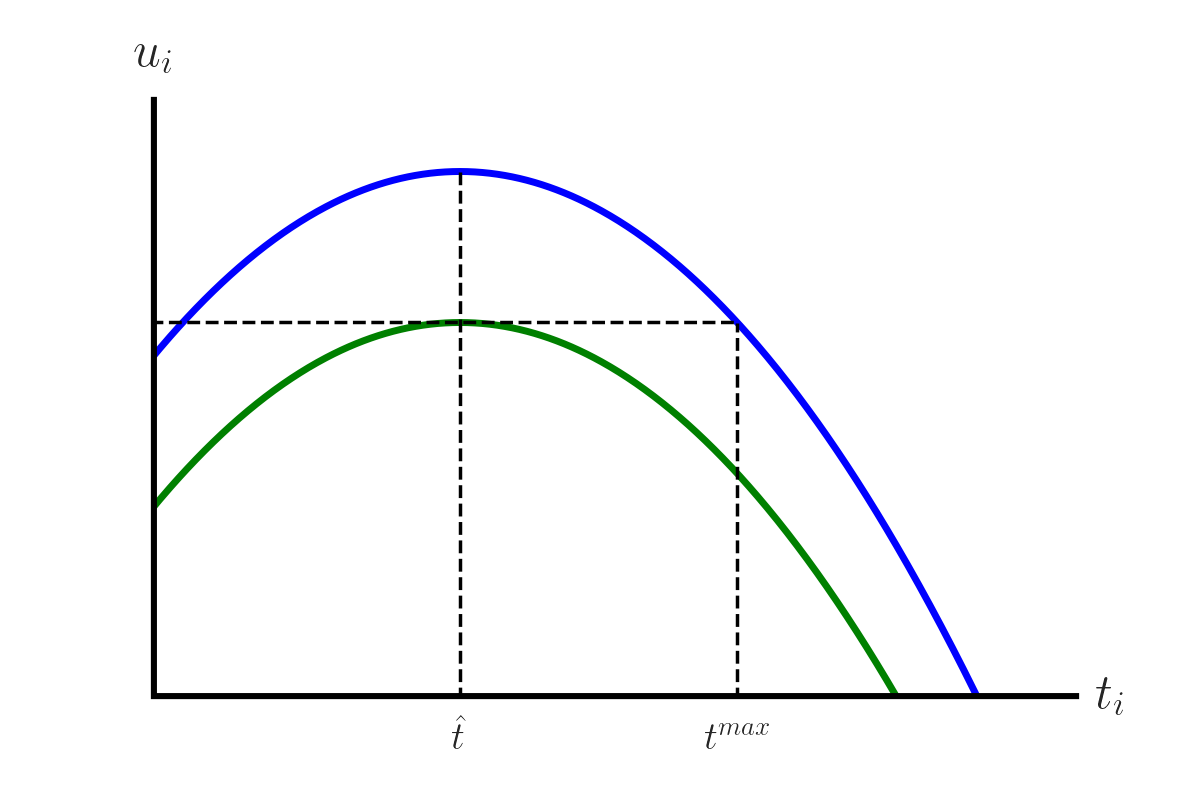} 
        \caption{}
        \label{fig:1B}
        \end{subfigure}
    \hfill
    \begin{subfigure}[b]{0.475\textwidth}  
        \centering 

        \includegraphics[width=\textwidth]{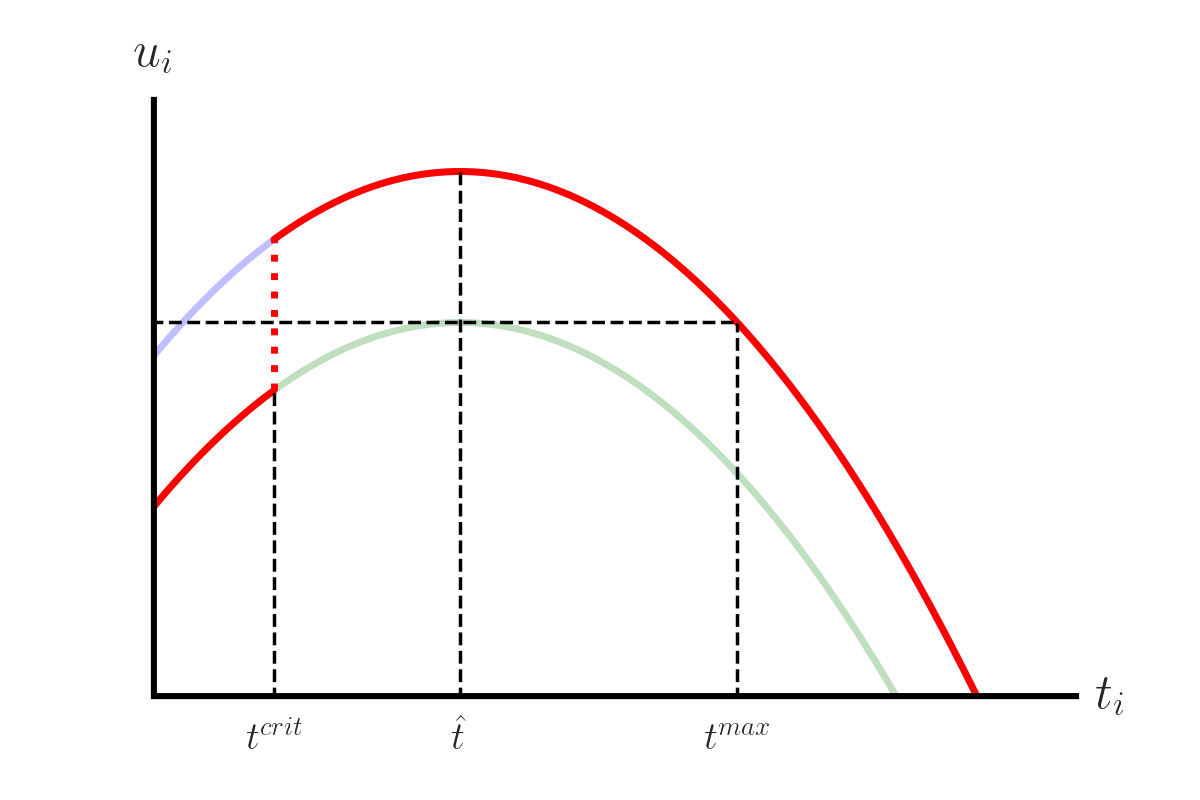}
        \caption{}
        \label{fig:2A}
    \end{subfigure}
    \vskip\baselineskip
    \begin{subfigure}[b]{0.475\textwidth}   
        \centering 
        \includegraphics[width=\textwidth]{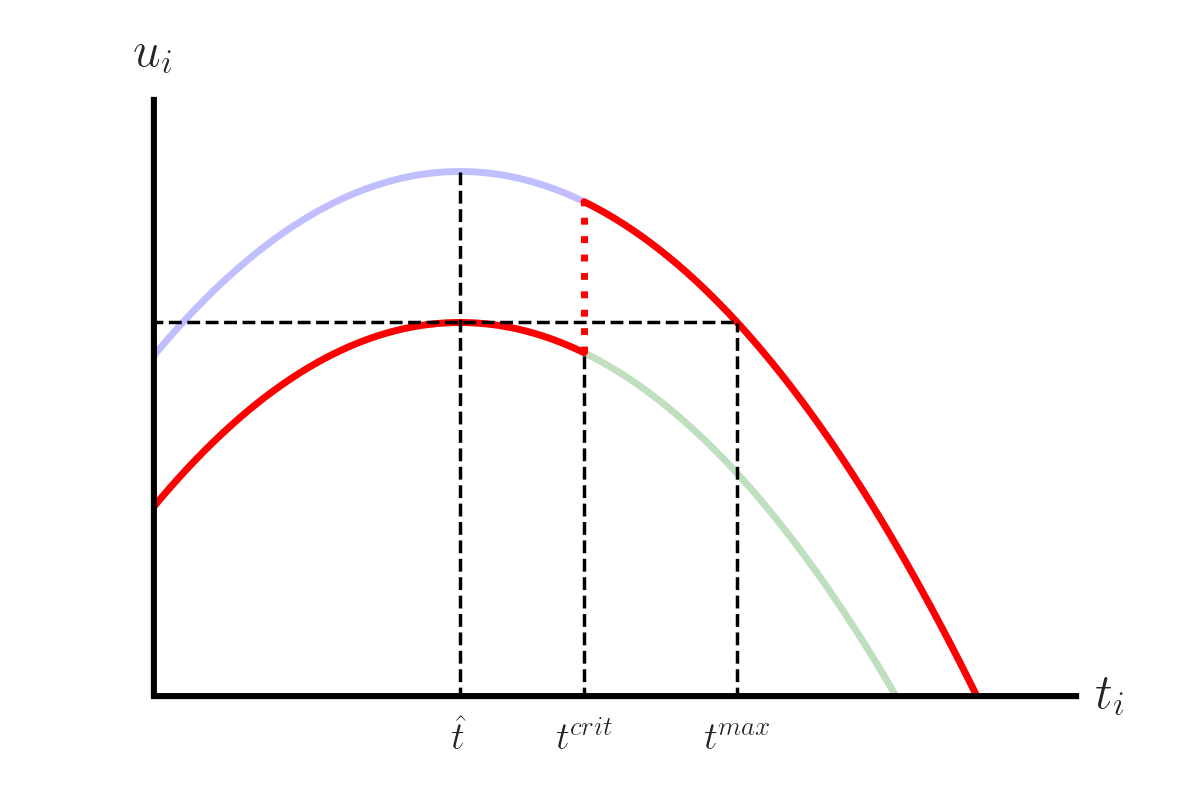}
        \caption{}
        \label{fig:2B}
    \end{subfigure}
    \hfill
    \begin{subfigure}[b]{0.475\textwidth}   
       \centering 
        \includegraphics[width=\textwidth]{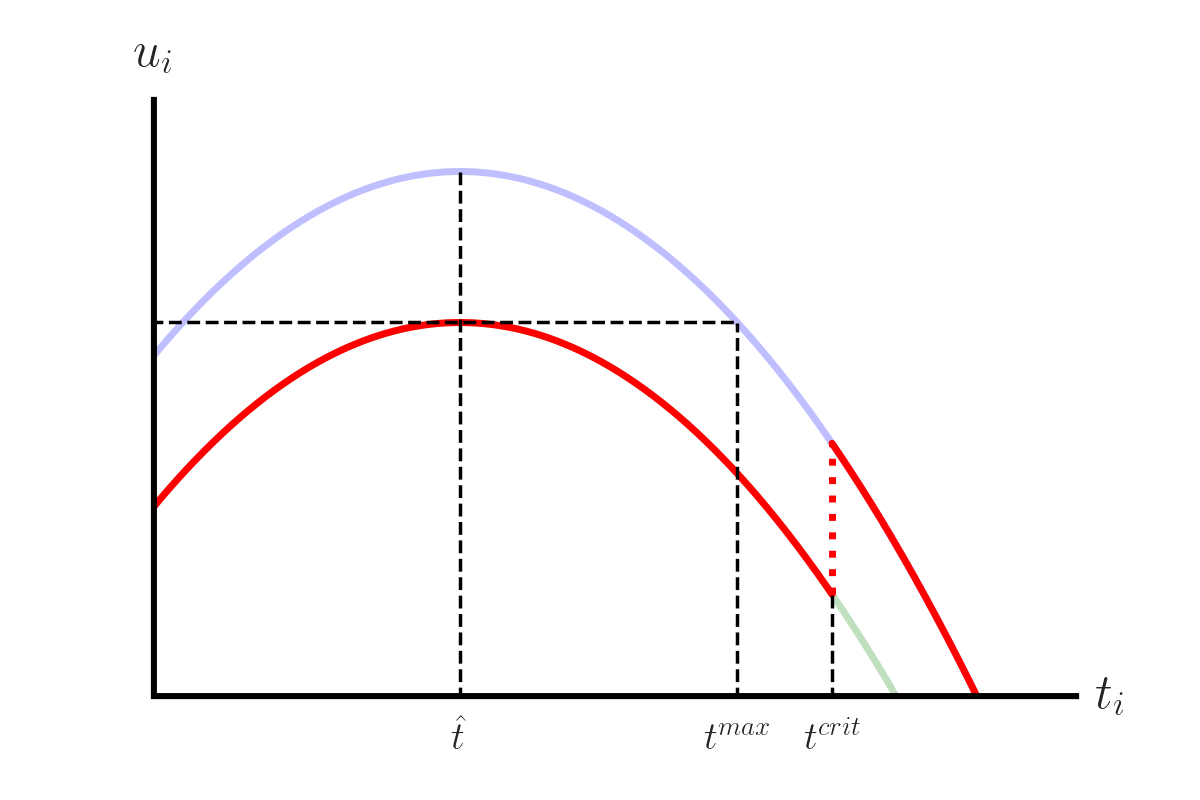}
        \caption{}
        \label{fig:2C}
    \end{subfigure}
    \caption{Blue line [resp. green line] shows payoffs with [resp. without] high public good provision, as a function of $t_h$, assuming optimal choice of $t_h^{out}$. Red line shows utility function assuming optimal choice of $t_h^{out}$. Panel (A): payoffs only.  Panel (B): $t^{crit}_h < \hat{t}$. Panel (C): $t^{crit}_h \in [\hat{t}, t^{max}]$. Panel (D): $t^{crit}_h > t^{max}$.}
    \label{fig:best_responses}
\end{figure}

From there, the final step is to show that heads of households simply consider whether it is worth deviating upwards from $\hat{t}_h$ (which is what they could guarantee if they ignored the impact they have on the second stage) to the point where they secure `high' public good provision in the second stage. \Cref{fig:best_responses} shows this final step graphically.

\subsection{Equilibrium}
Finding and characterising the equilibria of the game now boils down to finding fixed points of the best response function shown in \Cref{rem:best_response}.
To do this, I break up the possibile outcomes into four cases. Consider each in turn.


\paragraph{Case 1.} First, suppose that all households myopically maximising their first stage payoffs leads to the `high' second stage payoffs. That is, that all households choosing $t_h = \hat{t}_h$ induces network connectivity above the critical threshold (i.e. $t^{crit}( \ \hat{t}_{-h} ) \leq \hat{t}_h$ for all $h$). Then this is the unique equilibrium. 

Notice that if $t_h < \hat{t}_h$ for any $h$, then a household has a profitable deviation. This is because second stage payoffs are weakly increasing in $t_h$, no matter what. So if a household strictly increases its first stage payoffs by deviating, then it will always do so. So we must have $t_h \geq \hat{t}_h$ for any $h$. But once we are in this situation, second stage payoffs are `high' there are no further gains from more network connectivity. So any household with $t_h > \hat{t}_h$ has a profitable deviation -- to $t_h - \hat{t}_h$. So $t_h \neq \hat{t}_h$ for any $h$ is \emph{not} an equilibrium. It it then clear that $t_h = \hat{t}_h$ for all $h$ \emph{is} an equilibrium. A household's first stage payoffs strictly fall by deviating from this point. And their second stage payoffs weakly fall.

\paragraph{Case 2.} Second, suppose that all households choosing $\hat{t}_h$ would induce network connectivity below the threshold, but that network connectivity would cross the threshold if exactly one household changed their choice to some $t \leq t^{max}_h$. That is, if $\hat{t}_h < t^{crit}(\hat{t}_{-h})$ for all $h$ but that $t^{crit}(\hat{t}_{-h}) \leq t^{max}_h$ for some $h$.

Then households will always make first stage choices such that network connectivity is exactly at the critical threshold. This is because they will never choose below $\hat{t}_h$ (see argument above), and once there, at least one household has a profitable deviation to provide the additional network connectivity needed to restore the social fabric and obtain `high' second stage payoffs.

This equilibrium is not unique. There could be many different sets of choices $(t_h)_{h=1}^H$ that yield network connectivity exactly at the threshold but which involve no households choose above $t^{max}_h$. It is not possible to say which household (or households) provide the extra network connectivity to reach the threshold. Note that it could be the case that the households who end up choosing $t_h > \hat{t}_h$ could be ones who would not want to \emph{unilaterally} deviate from $t_h = \hat{t}_h$ for all $h$. I could be that 3 different households all choose $t_h > \hat{t}_h$, collectively bringing network connectivity up to the threshold, even though none of them would want to deviate by themselves. 

However, there is only one \emph{type} of equilibrium: households must make choices of $t_h$ that collectively induce network connectivity exactly at the threshold. 

\paragraph{Case 3.} Third, as above but that between $2$ and $H$ households must change their choice to some $t \in (\hat{t}_h , t^{max}_h)$ for network connectivity to rise to the critical threshold (and that no one household could get there alone). Here, there are \emph{two types} of equilibria. In one, agents coordinate on a Pareto efficient outcome, where they provide just enough network connectivity to maintain `high' public good contributions in the second stage. This type of equilibrium has the same structure as in Case 2. There are many equilibria for the same reasons. In the other, they mis-coordinate and follow their own first stage payoffs. This equilibrium is the same as the one in Case 1. But here, it is inefficient. Households are stuck there because no one household wants to deviate.

\paragraph{Case 4.}
Finally, suppose that all households choosing $t^{max}_h$ -- the maximum amount of time they would be willing to spend inside their community in order to preserve the social fabric -- \emph{still} does not induce network connectivity above the threshold. That is, $t^{max}_h < t^{crit}_h(t^{max}_{-h})$ for all $h$ (which in turn also implies that $\hat{t}_h < t^{crit}(\hat{t}_{-h})$ because we clearly have that $\hat{t}_h < t^{max}_h$ for all $h$, and $t^{crit}(\cdot)$ is a strictly decreasing function).

Once again, the equilibrium is unique. Households then simply maximise their first stage payoffs, and ignore the second stage. The argument is the same as in Case 1 except that now a deviation upward (to some $t_h > \hat{t}_h$) could increase the second stage payoffs. But such a deviation cannot be profitable -- essentially by assumption, as in Case 4 we have that $t^{max}_h < t^{crit}_h(t^{max}_{-h})$ for all $h$.

\begin{prop}\label{prop:eqm_formation} 
\emph{(Equilibrium)} 
In all convergent Nash equilibria, there exists an $n_0$ such that for all $n > n_0$:

\begin{itemize}
    \item[(i)] if $t^{crit}( \ \hat{t}_{-h} ) \leq \hat{t}_h$ for all $h$: then $t_h^* = \hat{t}_h$ for all $h$

    \item[(ii)] if $\hat{t}_h < t^{crit}(\hat{t}_{-h})$ for all $h$ and $t^{crit}(\hat{t}_{-h}) \leq t^{max}_h$ for some $h$: then $t_h^* = t^{crit}(t_{-h})$ and $t_h^* \in [\hat{t}_h, t^{max}_h]$ for all $h$ 

    \item[(iii)] if $t^{crit}(t^{max}_{-h}) \leq t^{max}_h < t^{crit}(\hat{t}_{-h} )$ for all $h$: then either (a)  $t_h^* = \hat{t}_h$ for all $h$ or (b) $t_h^* = t^{crit}(t_{-h})$ and $t_h^* \in [\hat{t}_h, t^{max}_h]$ for all $h$

    \item[(iv)] if $t^{max}_h < t^{crit}_h(t^{max}_{-h})$ for all $h$: then $t_h^* = \hat{t}_h$ for all $h$.
\end{itemize}
\end{prop}

This result follows from the discussion preceding it, plus \Cref{rem:best_response}. 
Comparative statics remain straightforward. $\hat{t}_h$ is still decreasing in $\pi_h$ -- its definition has not changed compared to the main model (only notice that $\hat{t}_h$ is the value of $\hat{t}_i$ for the agent $i$ who is the `head' of household $h$). $t^{max}_h$ is also decreasing in $\pi_h$. An increase in $\pi_h$ raises the opportunity cost of spending time inside the community in all circumstances. So the maximum amount of time a household would be willing to spend inside the community to access `high' second stage payoffs must fall.

\begin{rem}\label{prop:comp_stat_formation} 
$\hat{t}_h$ and $t^{max}_h$ are strictly decreasing in $\pi_h$, for all $h$.
\end{rem}

The presence of multiple equilibria makes it more difficult to make unambiguous statements about welfare. However, it is straightforward to see that when multiple equilibria exist, all agents prefer the type where they spend just enough time inside the community to maintain the social fabric (i.e. they locate exactly at the critical threshold) over the type where they simply choose $\hat{t}_h$. 

This is essentially true by design -- any household could revert to playing $\hat{t}_h$ and lose the `high' second stage payoffs. Once these `high' second stage payoffs are lost, no other household's choices can reduce $h$'s payoffs. So if locating exactly at the critical threshold \emph{is} an equilibrium, then all households must prefer it over reverting to playing $\hat{t}_h$. 

But note that there are many (asymmetric) equilibria with households collectively spending just enough time inside the community to locate exactly at the critical threshold. The exact amount of time each household spends inside the community is not pinned down. And it is not generally possible to provide a welfare ranking of these equilibria. Conditional on being at the critical threshold, each household wants to spend less time inside the community (down to a minimum of $\hat{t}_h$).

\paragraph{Fragility.} These equilibria where agents locate exactly at the critical threshold are fragile in the sense that arbitrarily small unanticipated shocks can suddenly tear the social fabric. This is precisely because the no household wants to provide more network connectivity than is needed to \emph{just} maintain the social fabric. These `fragile' equilibria are not a knife-edge case. In fact they exist whenever uncoordinated decisions lead to low second stage payoffs (i.e. no social fabric), but coordination amongst \emph{all} agents/households in the community would lead to high second stage payoffs (i.e. a functioning social fabric).

However, these same equilibria are \emph{robust} to very small \emph{anticipated} shocks. In fact, the shock I added in \Cref{sec:model_extended} is precisely a small anticipated shock. This is because for small anticipated shocks, households understand that increasing network connectivity a little bit will, given the coming shock, increase the probability that network connectivity post-shock remains above the critical threshold. When the shock is small, it is worth `paying for' some extra network connectivity (in the form of lower first stage payoffs) in order to increase this probability.

\newpage
\subsection{Proof of \Cref{rem:best_response}}

\paragraph{Step 1: express network connectivity as a function of $t_h$ and $\xi$.} By definition, we have $\tilde{d}_i = d_i$ for all agents not affected by the shock. For those affected by the shock we have $\tilde{d}_i = d_i + 1$ if $\xi > 0$ and $\tilde{d}_i = d_i - 1$ if $\xi < 0$. Define $\Xi := \{i : i \text{ affected by shock}\}$. And by definition, $n^{-1} \cdot |\Xi| = \xi$. Therefore when $\xi > 0$:

\begin{align}
    \lim_{n \to \infty} \frac{1}{n} \sum_i \tilde{d}_i (\tilde{d}_i - 2) 
    &= \lim_{n \to \infty} \frac{1}{n} \sum_{i \ \notin \ \Xi} d_i (d_i - 2) + \lim_{n \to \infty} \sum_{i \ \in \ \Xi} (d_i + 1) (d_i - 1) \\
    &= \lim_{n \to \infty} \frac{1}{n} \sum_{i = 1}^n d_i (d_i - 2) + \xi (2 \bar{d} - 1) 
\end{align}
where $\bar{d} := \frac{1}{|\Xi|} \sum_{i \in \Xi} d_i$. And when $\xi < 0$:
\begin{align}
    \lim_{n \to \infty} \frac{1}{n} \sum_i \tilde{d}_i (\tilde{d}_i - 2) 
    &= \lim_{n \to \infty} \frac{1}{n} \sum_{i \ \notin \ \Xi} d_i (d_i - 2) + \lim_{n \to \infty} \sum_{i \ \in \ \Xi} (d_i - 1) (d_i - 3) \\
    &= \lim_{n \to \infty} \frac{1}{n} \sum_{i = 1}^n d_i (d_i - 2) - \xi (2 \bar{d} - 3) 
\end{align}
Note that our assumption that the shock only hits people who have $d_i \geq 2$ means that $(2 \bar{d} - 1) > (2 \bar{d} - 3) > 0$. So let $k = (2 \bar{d} - 1) > 0$ if $\xi >0$ and $k= (2 \bar{d} - 3) > 0$ if $\xi < 0$. Then we know from \Cref{lem:threshold_t_hat} that $\lim_{n \to \infty} \frac{1}{n} \sum_i d_i (d_i - 2) = \sum_h f_h Z(t_h)$ for some strictly increasing function $Z(\cdot)$. And here, we have assumed that all households are equal-sized. So $f_h = \frac{1}{H}$ for all $H$. Hence we have
\begin{align}
    \lim_{n \to \infty} \frac{1}{n} \sum_i \tilde{d}_i (\tilde{d}_i - 2) = \frac{1}{H} \sum_h Z(t_h) + k \xi
\end{align}

\paragraph{Step 2: show that households never choose a level of network connectivity where the realisation of $\xi$ matters.} Start by defining
\begin{align}
    P_{SC} := Pr \left( \lim_{n \to \infty} \frac{1}{n} \sum_{i=1}^n \tilde{d}_i (\tilde{d_i} - 2) > 0 \right) = Pr\left( \frac{1}{H} \sum_h Z(t_h) + k \xi > 0 \right)
\end{align}
Then rearranging terms and applying the CDF of $\xi$ in the standard way yields: 
\begin{align}
    P_{SC} = 1 - \widehat{F}\left(- \frac{1}{\bar{\xi}} \frac{1}{H} \frac{1}{k} \sum_h Z(t_h)  \right).
\end{align}
Now by the definition of $\hat{F}(\cdot)$, we have $P_{SC} = 1$ whenever
\begin{align}
    \frac{1}{H} \frac{1}{k} \sum_h Z(t_h) &\geq \bar{\xi}.
\end{align}
Consider some household $h$ and fix $t_{h'}$ for all $h' \neq h$. Then we have $P_{SC} = 1$ whenever
\begin{align}\label{eq:t_crit_defn}
    Z(t_{h}) \geq \bar{\xi} H k - \sum_{h' \neq h} Z(t_{h'})
\end{align}

Since $Z(t_h)$ is a strictly increasing function, and $Z(t_h) = 0$ for some $t_h \in (0,1)$, then there must be a unique value of $t_h$ that satisfies \cref{eq:t_crit_defn} with equality. Denote this value $t_h^{crit}$. It is a function of all other $t_h$'s; $t_h^{crit} = t^{crit}(t_{-h})$. Note however, that there is absolutely no guarantee that this value is feasible (i.e. it is on the $[0,1]$ interval and so is something that household $h$ could actually choose). 
Therefore, there is a unique $t^{crit}_h = t^{crit}(t_{-h})$ such that $P_{SC} = 1$ for all $t_h \geq t^{crit}_h$. \textbf{Call this observation 1.}

Similarly, we have $P_{SC} = 0$ whenever $\frac{1}{H} \frac{1}{k} \sum_h Z(t_h) \leq - \bar{\xi}$. So by identical logic to immediately above, there is a unique $t^{uncrit}_h = t^{uncrit}(t_{-h})$ such that $P_{SC} = 0$ for all $t_h \leq t^{uncrit}_h$. \textbf{Call this observation 2.} 

In-between these two bounds ($t^{uncrit}_h$ and $t^{crit}_h$), the probability of being above the threshold, $P_{SC}$, is strictly increasing in $t_h$. 
More precisely, we have:
\begin{align}
    \frac{d P_{SC}}{d t_h} = \widehat{F}'(\cdot) \frac{1}{\bar{\xi}} \frac{1}{H} \frac{1}{k} Z'(t_h) \text{ 
 when } t_{h} \in (t^{uncrit}_h, t^{crit}_h).
\end{align}

Therefore, when $t_{h} \in (t^{uncrit}_h, t^{crit}_h)$, we have $\frac{d P_{SC}}{d t_h} \to \infty$ as $\bar{\xi} \to 0$. Finally, notice that the difference between the two bounds ($t^{uncrit}_h$ and $t^{crit}_h$) shrinks to zero as $\bar{\xi} \to 0$. 

From observation 1, we have that $Z(t_h^{crit}) = \bar{\xi} H k - \sum_{h' \neq h} Z(t_{h'})$. And from observation 2, we have that $Z(t_h^{uncrit}) = -\bar{\xi} H k - \sum_{h' \neq h} Z(t_{h'})$. Therefore, we have $Z(t_h^{crit}) - Z(t_h^{uncrit}) = 2 \bar{\xi} H k$. Finally, as $\bar{\xi} \to 0$, we must have $Z(t_h^{crit}) - Z(t_h^{uncrit}) \to 0$ and hence that $t_h^{crit} - t_h^{uncrit} \to 0^+$.

\paragraph{Step 3: show that households choose $\hat{t}_h$ or $t^{crit}$.} We can now write the utility function as $u_i = w_h(t_h) + P_{SC} \Delta v + const.$ Taking $\lim \bar{\xi} \to 0$, we then have:
\begin{align}\label{eq:utility_derivative}
    \frac{d \lim_{\bar{\xi} \to 0} u_i}{d t_h} = 
    \begin{cases}
        \infty &\text{ for } t_h \in (t^{uncrit}, t^{crit}), \\
        \frac{d w_h(t_h)}{d t_h} &\text{ otherwise. }   
    \end{cases}
\end{align}
Therefore the utility function is as shown in \Cref{fig:best_responses} (panels (B)-(D)), and best responses follow from inspection of the figure. \hfill \qed 

\end{document}